\documentclass[journal]{IEEEtran}

\usepackage{epsfig,amsmath,amssymb,epsf,amsthm,scalefnt,multirow,bm,mathtools,stfloats,stmaryrd}
\usepackage{xcolor}
\usepackage{float}
\usepackage{cite}
\usepackage{psfrag}
\usepackage{algorithm}
\usepackage{algpseudocode}

\newtheorem*{lemma*}{Lemma}

\algrenewcommand\algorithmicrequire{\textbf{Input:}}
\algrenewcommand\algorithmicensure{\textbf{Output:}}
\algdef{SE}[DOWHILE]{Do}{doWhile}{\algorithmicdo}[1]{\algorithmicwhile\ #1}

\begin{document}

\title{Bayesian Channel Estimation for Intelligent Reflecting Surface-Aided mmWave Massive MIMO Systems With Semi-Passive Elements}

\author{In-soo Kim, Mehdi Bennis, Jaeky Oh, Jaehoon Chung, and Junil Choi
\thanks{I. Kim is with Wireless Research \& Development (WRD), Qualcomm Technologies, Inc., San Diego, CA, USA (e-mail: insookim@qti.qualcomm.com).}
\thanks{J. Choi is with the School of Electrical Engineering, KAIST, Daejeon, South Korea (e-mail: junil@kaist.ac.kr).}
\thanks{M. Bennis is with the Centre for Wireless Communications, University of Oulu, Oulu, Finland (e-mail: mehdi.bennis@oulu.fi).}
\thanks{Jaeky Oh and Jaehoon Chung are with C\&M Standard Lab, ICT Technology Center, LG Electronics Inc. (e-mail: \{jaeky.oh; jaehoon.chung\}@lge.com).}
\thanks{This work was supported in part by LG Electronics Inc., in part by the MSIT (Ministry of Science and ICT), Korea, under the ITRC (Information Technology Research Center) support program (IITP-2020-0-01787) supervised by the IITP (Institute of Information \& Communications Technology Planning \& Evaluation), in part by the Ministry of Trade, Industry and Energy (MOTIE) and Korea Institute for Advancement of Technology (KIAT) through the International Cooperative R\&D program (P0022557), and in part by the Institute of Information \& Communications Technology Planning \& Evaluation (IITP) grant funded by the Korea government (MSIT) (No. 2021-0-00269, Development of sub-THz band wireless transmission and access core technology for 6G Tbps data rate).}}

\maketitle

\begin{abstract}
In this paper, we propose a Bayesian channel estimator for intelligent reflecting surface-aided (IRS-aided) millimeter wave (mmWave) massive multiple-input multiple-output (MIMO) systems with semi-passive elements that can receive the signal in the active sensing mode. Ultimately, our goal is to minimize the channel estimation error using the received signal at the base station and additional information acquired from a small number of active sensors at the IRS. Unlike recent works on channel estimation with semi-passive elements that require both uplink and downlink training signals to estimate the UE-IRS and IRS-BS links, we only use uplink training signals to estimate all the links. To compute the minimum mean squared error (MMSE) estimates of all the links, we propose a novel variational inference-sparse Bayesian learning (VI-SBL) channel estimator that performs approximate posterior inference on the channel using VI with the mean-field approximation under the SBL framework. The simulation results show that VI-SBL outperforms the state-of-the-art baselines for IRS with passive reflecting elements in terms of the channel estimation accuracy and training overhead. Furthermore, VI-SBL with semi-passive elements is shown to be more spectral- and energy-efficient than the baselines with passive reflecting elements.
\end{abstract}


\section{Introduction}
\IEEEPARstart{5}{G} wireless communications support high data rates by communicating in the millimeter wave (mmWave) band \cite{8373698}. The high carrier frequency in the range of 30-300 GHz offers a large bandwidth, which results in a significant throughput gain. The problem, however, is that the severe path loss renders mmWave communications vulnerable to blockages. To overcome such an issue, intelligent reflecting surface (IRS) was recently proposed \cite{8910627}, which is an array of metamaterial-based passive reflecting elements capable of adjusting the amplitude and phase of the impinging signal. For the IRS to generate a favorable detour around a blockage, the reflection amplitude and phase shift must align with the channel, which necessitates accurate channel state information (CSI).

The distinct feature of channel estimation for IRS with passive reflecting elements is that the UE-IRS link of size $NK$ and IRS-BS link of size $MN$ form a UE-IRS-BS link of size $MNK$ where $M$, $N$, and $K$ are the numbers of base station antennas, IRS elements, and single-antenna users. Since passive reflecting elements cannot observe the UE-IRS and IRS-BS links, the UE-IRS-BS link must be estimated from the reflected signal, which results in a significant training overhead.

Next, we review prior works \cite{9130088, 9133142, 9195133, 9133156, 9400843, 9398559, 9707728, 9103231, 9328485} on channel estimation for IRS with passive reflecting elements that focus on training overhead reduction. In \cite{9130088}, a three-phase channel estimator is proposed inspired by the correlation between the channels of different users. In particular, \cite{9130088} exploits the fact that the UE-IRS-BS links of different users share the same IRS-BS link. As a result, the three-phase channel estimator proceeds by estimating the UE-BS links of all the users in the first phase, UE-IRS-BS link of a particular user in the second phase, and UE-IRS-BS links of the remaining users in the third phase where most of the training overhead reduction occurs. Moving on to \cite{9133142}, channel estimation for IRS with discrete phase shifts is considered, for which a reflection design is proposed. In particular, the reflection design is optimized based on grouping IRS elements to reduce the channel estimation error. In \cite{9195133}, a channel estimator is proposed for the orthogonal frequency division multiple access (OFDMA) scenario based on the line-of-sight (LoS) assumption on the UE-BS link. Again, the channel estimator exploits the fact that the UE-IRS-BS links of different users share the same IRS-BS link to reduce the training overhead as in \cite{9130088}. In \cite{9133156}, the channel estimation problem is reformulated as a matrix factorization problem, which is solved using the message passing (MP) algorithm. In \cite{9400843}, a dual link training signal-based channel estimator is proposed. In particular, the dual link training signal transmits downlink training signals to the IRS, whose reflected version is used as uplink training signals to estimate the IRS-BS link. In \cite{9398559}, an atomic norm minimization-based channel estimator is proposed that extracts the angle parameters of the channel. In \cite{9707728}, a channel estimator is proposed based on the single-path approximation of the channel. In addition, \cite{9707728} develops another channel estimator that exploits IRS phase shifts and training signals. In \cite{9103231}, the mmWave channel estimation is reformulated as a compressed sensing problem to capture the channel sparsity. In \cite{9328485}, the double-structured orthogonal matching pursuit (DS-OMP) channel estimator is proposed that exploits the common IRS-BS link shared by all the users. In particular, DS-OMP reduces the channel estimation error and training overhead by identifying the common rows and columns that the UE-IRS-BS links of all the users share by taking into account the double sparsity structure inherent in the IRS links.

The training overhead in \cite{9130088, 9133142, 9195133, 9133156, 9400843, 9398559, 9707728, 9103231, 9328485} is still high because estimating the UE-IRS-BS link with passive reflecting elements that cannot receive the signal is a challenging task. To overcome such an issue, IRS with semi-passive elements that can be switched to the active sensing mode was recently proposed in \cite{9511813, 9529045}. In particular, the received signal at the active sensors is leveraged to estimate the UE-IRS and IRS-BS links in the first coherence block. Then, \cite{9511813, 9529045} can replace UE-IRS-BS link estimation of size $MNK$ with UE-IRS link estimation of size $NK$ in the subsequent coherence blocks because the IRS-BS link remains constant over multiple coherence blocks, which results in significant channel estimation overhead reduction. The problem of \cite{9511813, 9529045}, however, is that both uplink and downlink training signals are necessary to estimate all the links. Furthermore, \cite{9511813, 9529045} are semi-passive element-\textit{driven} rather than \textit{aided} because only the received signal at the active sensors is leveraged, while the received signal at the base station is discarded.

In this paper, a Bayesian channel estimator is proposed for IRS-aided mmWave massive multiple-input multiple-output (MIMO) systems with semi-passive elements. In particular, the semi-passive elements are activated to the active sensing mode in the channel estimation phase and deactivated to the passive reflecting mode in the data transmission phase. For the first time in the literature, we estimate the UE-IRS and IRS-BS links using only uplink training signals, which is in contrast to recent works \cite{9511813, 9529045} on channel estimation with semi-passive elements that rely on both uplink and downlink training signals to estimate all the links. To compute the minimum mean squared error (MMSE) estimate of the channel from the received signal at the base station and additional information acquired from the active sensors at the IRS, we perform posterior inference on the channel under the sparse Bayesian learning (SBL) framework \cite{10.1162/15324430152748236, 1315936}. Since exact posterior inference is intractable, we use the variational inference (VI) approach with the mean-field approximation \cite{4644060, MAL-001}. The proposed VI-SBL-based channel estimator enables us to compute the approximate MMSE estimates of all the links iteratively. In addition, we reduce the complexity of VI-SBL by converting a large matrix inversion to many small matrix inversions using the mean-field approximation. The simulation results show that VI-SBL outperforms the state-of-the-art channel estimators for IRS with passive reflecting elements in terms of the channel estimation error, training overhead, and energy efficiency, which is defined as the spectral efficiency normalized by the total power consumption to make a fair comparison between IRS with semi-passive elements and passive reflecting elements.

The rest of the paper is organized as follows. The channel model and signal model are introduced in Section \ref{section_2}. In Section \ref{section_3}, the proposed VI-SBL-based channel estimator is developed, which is followed by the mean-field approximation-based complexity reduction scheme. The performance of VI-SBL is assessed in Section \ref{section_4} based on various performance metrics, and a concluding remark follows in Section \ref{section_5}.

\textbf{Notation:} $a$, $\mathbf{a}$, and $\mathbf{A}$ denote a scalar, vector, and matrix. The complex conjugate, transpose, and conjugate transpose of $\mathbf{A}$ are written as $\mathbf{A}^{*}$, $\mathbf{A}^{\mathrm{T}}$, and $\mathbf{A}^{\mathrm{H}}$. $|a|$ and $\angle a$ are the magnitude and phase of $a$. The $i$-th element of $\mathbf{a}$ is $a_{i}$, while the $(i, j)$-th element and $i$-th column of $\mathbf{A}$ are $[\mathbf{A}]_{i, j}$ and $[\mathbf{A}]_{:, i}$. The elementwise product, Kronecker product, and Khatri-Rao product of $\mathbf{A}$ and $\mathbf{B}$ are written as $\mathbf{A}\odot\mathbf{B}$, $\mathbf{A}\otimes\mathbf{B}$, and $\mathrm{kr}(\mathbf{A}, \mathbf{B})$. The $n\times 1$ all-zero vector, $n\times 1$ all-one vector, and $n\times n$ identity matrix are written as $\mathbf{0}_{n}$, $\mathbf{1}_{n}$, and $\mathbf{I}_{n}$. $\mathrm{vec}(\mathbf{A})$ is the vectorization of $\mathbf{A}$, while $\mathrm{diag}(\mathbf{a})$ is the diagonal matrix with $\mathbf{a}$ on the main diagonal. The probability density function (PDF) of a complex Gaussian random vector $\mathbf{x}\sim\mathcal{CN}(\mathbf{m}, \mathbf{C})$ with mean $\mathbf{m}$ and covariance $\mathbf{C}$ is written as $\mathcal{CN}(\mathbf{x}|\mathbf{m}, \mathbf{C})$. The probability measure of a random vector $\mathbf{x}$ is $p(\mathbf{x})$. The set difference between sets $\mathcal{A}$ and $\mathcal{B}$ is $\mathcal{A}\setminus\mathcal{B}$. $\llbracket N\rrbracket$ denotes $\{1, ..., N\}$.

\begin{figure}[t]
\centering
\includegraphics[width=1\columnwidth]{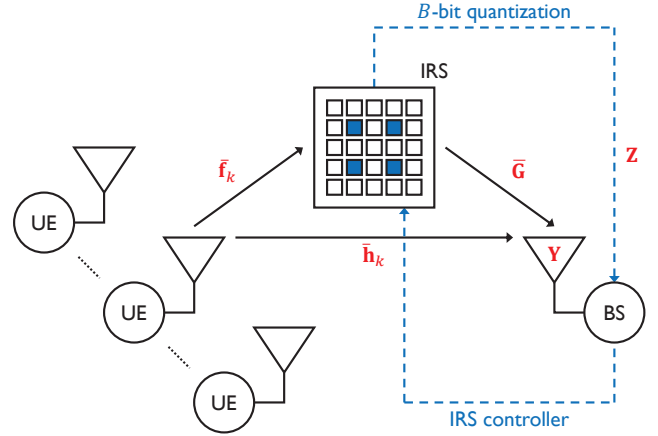}
\caption[caption]{The uplink of an IRS-aided mmWave massive MIMO system with semi-passive elements. The semi-passive elements act as active sensors in the channel estimation phase and passive reflecting elements in the data transmission phase.}\label{figure_1}
\end{figure}

\section{System Model and Problem Formulation}\label{section_2}
Consider the uplink of an IRS-aided mmWave massive MIMO system with an $M$-antenna base station and $K$ single-antenna users as illustrated in Fig. \ref{figure_1}. The IRS is equipped with $N$ elements to compensate for the path loss in the mmWave band. In particular, $N_{\mathrm{p}}$ passive reflecting elements reflect the impinging signal as usual \cite{8811733, 9115725, 9326394}. Meanwhile, $N_{\mathrm{a}}=N-N_{\mathrm{p}}\ll N$ semi-passive elements act as active sensors in the channel estimation phase and passive reflecting elements in the data transmission phase. The active sensors are capable of receiving the signal, whose quantized version is forwarded to the base station. In practice, the active sensors are implemented by connecting the IRS elements to $N_{\mathrm{a}}$ radio frequency (RF) chains with $B$-bit analog-to-digital converters (ADCs) \cite{9511813}, whose connection can be switched using base station-controlled switching networks \cite{7370753, 7880698}. In this paper, a small number of active sensors and low-resolution ADCs are considered to reduce the additional power consumed by the IRS.

In essence, our goal is to exploit the received signal at the base station and additional information acquired from the active sensors at the IRS to estimate the UE-IRS and IRS-BS links instead of the UE-IRS-BS link. Then, the base station can avoid estimating the UE-IRS-BS link of size $MNK$ and only estimate the UE-IRS link of size $NK$ in the subsequent coherence blocks because the IRS-BS link remains constant over multiple coherence blocks in practice \cite{9400843}. As a result, channel estimation overhead reduction in the long run is attained from the reduced training overhead achieved by the active sensors in the first coherence block, and replacing UE-IRS-BS link estimation with UE-IRS link estimation in the subsequent coherence blocks. In this paper, we focus on the first coherence block where all the links are unknown.

\subsection{Channel Model}
The channel is composed of the UE-IRS link $\bar{\mathbf{f}}_{k}\in\mathbb{C}^{N}$ for $k\in\llbracket K\rrbracket$, IRS-BS link $\bar{\mathbf{G}}\in\mathbb{C}^{M\times N}$, and UE-BS link $\bar{\mathbf{h}}_{k}\in\mathbb{C}^{M}$ for $k\in\llbracket K\rrbracket$. Since the scatterers are limited in the mmWave band as the path loss is severe, the links are typically modeled as \cite{6834753}
\begin{align}
\bar{\mathbf{f}}_{k}=&\sqrt{\frac{N\kappa_{\mathrm{UI}, k}}{1+\kappa_{\mathrm{UI}, k}}}\alpha_{\mathrm{UI}, k, 0}\mathbf{a}_{\mathrm{I}}(\theta_{\mathrm{UI}, k, 0}^{\mathrm{AoA}}, \theta_{\mathrm{UI}, k, 0}^{\mathrm{ZoA}})+\notag\\
                     &\sqrt{\frac{N}{L_{\mathrm{UI}, k}(1+\kappa_{\mathrm{UI}, k})}}\sum_{\ell=1}^{L_{\mathrm{UI}, k}}\alpha_{\mathrm{UI}, k, \ell}\mathbf{a}_{\mathrm{I}}(\theta_{\mathrm{UI}, k, \ell}^{\mathrm{AoA}}, \theta_{\mathrm{UI}, k, \ell}^{\mathrm{ZoA}}),\notag\\
    \bar{\mathbf{G}}=&\sqrt{\frac{MN\kappa_{\mathrm{IB}}}{1+\kappa_{\mathrm{IB}}}}\alpha_{\mathrm{IB}, 0}\mathbf{a}_{\mathrm{B}}(\theta_{\mathrm{IB}, 0}^{\mathrm{AoA}})\mathbf{a}_{\mathrm{I}}^{\mathrm{H}}(\theta_{\mathrm{IB}, 0}^{\mathrm{AoD}}, \theta_{\mathrm{IB}, 0}^{\mathrm{ZoD}})+\notag\\
                     &\sqrt{\frac{MN}{L_{\mathrm{IB}}(1+\kappa_{\mathrm{IB}})}}\sum_{\ell=1}^{L_{\mathrm{IB}}}\alpha_{\mathrm{IB}, \ell}\mathbf{a}_{\mathrm{B}}(\theta_{\mathrm{IB}, \ell}^{\mathrm{AoA}})\mathbf{a}_{\mathrm{I}}^{\mathrm{H}}(\theta_{\mathrm{IB}, \ell}^{\mathrm{AoD}}, \theta_{\mathrm{IB}, \ell}^{\mathrm{ZoD}}),\notag\\
\bar{\mathbf{h}}_{k}=&\sqrt{\frac{M\kappa_{\mathrm{UB}, k}}{1+\kappa_{\mathrm{UB}, k}}}\alpha_{\mathrm{UB}, k, 0}\mathbf{a}_{\mathrm{B}}(\theta_{\mathrm{UB}, k, 0}^{\mathrm{AoA}})+\notag\\
                     &\sqrt{\frac{M}{L_{\mathrm{UB}, k}(1+\kappa_{\mathrm{UB}, k})}}\sum_{\ell=1}^{L_{\mathrm{UB}, k}}\alpha_{\mathrm{UB}, k, \ell}\mathbf{a}_{\mathrm{B}}(\theta_{\mathrm{UB}, k, \ell}^{\mathrm{AoA}})\label{fgh}
\end{align}
where a uniform linear array (ULA) geometry at the base station and uniform planar array (UPA) geometry at the IRS with half-wavelength spacings are assumed without loss of generality. For each link, $\alpha_{\ell}\sim\mathcal{CN}(0, 1/\mathrm{PL})$ is the $\ell$-th path gain where $\mathrm{PL}$ is the distance- and frequency-dependent path loss, $\theta_{\ell}$ is the $\ell$-th azimuth/zenith angle of arrival/departure, $\kappa$ is the Rician K-factor, and $L$ is the number of non-LoS (NLoS) paths. In addition, $\mathbf{a}_{\mathrm{B}}(\cdot)\in\mathbb{C}^{M}$ and $\mathbf{a}_{\mathrm{I}}(\cdot, \cdot)\in\mathbb{C}^{N}$ are the array response vectors of the base station and IRS.

\subsection{Signal Model}
Let $\bm{\Omega}[t]\in\{0, 1\}^{N}$ and $\bm{\Omega}^{\mathrm{c}}[t]=\mathbf{1}_{N}-\bm{\Omega}[t]$ denote the index vectors of the active sensors and passive reflecting elements that constitute the IRS at time $t$. In addition, define $\bar{\mathbf{F}}=[\bar{\mathbf{f}}_{1}, \dots, \bar{\mathbf{f}}_{K}]$ and $\bar{\mathbf{H}}=[\bar{\mathbf{h}}_{1}, \dots, \bar{\mathbf{h}}_{K}]$ for notational simplicity.

Then, the received signal $\mathbf{y}[t]\in\mathbb{C}^{M}$ at the base station is
\begin{equation}
\mathbf{y}[t]=\bar{\mathbf{G}}\left(\underbrace{\bm{\Omega}^{\mathrm{c}}[t]\odot\mathbf{v}[t]}_{=\mathbf{s}[t]}\odot\bar{\mathbf{F}}\mathbf{x}[t]\right)+\bar{\mathbf{H}}\mathbf{x}[t]+\mathbf{n}_{\mathrm{B}}[t]
\end{equation}
where $\mathbf{v}[t]\in\mathbb{C}^{N}$ is the passive reflection vector with the reflection amplitude $|v_{n}[t]|\leq 1$ and phase shift $\angle v_{n}[t]\in[0, 2\pi)$, $\mathbf{x}[t]\in\mathbb{C}^{K}$ is the transmit signal of the users under the transmit power constraint $\mathbb{E}\{|x_{k}[t]|^{2}\}\leq P_{k}[t]$, and $\mathbf{n}_{\mathrm{B}}[t]\sim\mathcal{CN}(\mathbf{0}_{M}, \sigma_{\mathrm{B}}^{2}\mathbf{I}_{M})$ is the additive white Gaussian noise (AWGN) at the base station. Likewise, the quantized received signal $\mathbf{z}[t]\in\mathbb{C}^{N}$ at the active sensors forwarded to the base station is
\begin{equation}
\mathbf{z}[t]=\bm{\Omega}[t]\odot\mathrm{Q}(\bar{\mathbf{F}}\mathbf{x}[t]+\mathbf{n}_{\mathrm{I}}[t])
\end{equation}
where $\mathbf{n}_{\mathrm{I}}[t]\sim\mathcal{CN}(\mathbf{0}_{N}, \sigma_{\mathrm{I}}^{2}\mathbf{I}_{N})$ is the AWGN at the active sensors. The $B$-bit quantizer $\mathrm{Q}(\cdot)$ is applied to the real and imaginary parts elementwise as
\begin{equation}\label{qu}
z=\mathrm{Q}(u)\iff\begin{cases}\mathrm{Re}(z^{\mathrm{lo}})\leq\mathrm{Re}(u)<\mathrm{Re}(z^{\mathrm{up}})\\\mathrm{Im}(z^{\mathrm{lo}})\leq\mathrm{Im}(u)<\mathrm{Im}(z^{\mathrm{up}})\end{cases}
\end{equation}
where $z^{\mathrm{lo}}\in\mathbb{C}$ and $z^{\mathrm{up}}\in\mathbb{C}$ are the lower and upper thresholds associated with $z\in\mathbb{C}$. In other words, the real and imaginary parts of $z$, $z^{\mathrm{lo}}$, and $z^{\mathrm{up}}$ correspond to one of the $2^{B}$ quantization intervals.

A coherence block of length $T_{\mathrm{c}}$ in the uplink consists of the channel estimation phase of length $T$ and data transmission phase of length $T_{\mathrm{c}}-T$. To proceed, let $\mathcal{T}_{\mathrm{c}}=\llbracket T\rrbracket$ and $\mathcal{T}_{\mathrm{d}}=\llbracket T_{\mathrm{c}}\rrbracket\setminus\llbracket T\rrbracket$ be the time slots for the channel estimation phase and data transmission phase. Then, the received signals $\mathbf{Y}\in\mathbb{C}^{M\times T}$ and $\mathbf{Z}\in\mathbb{C}^{N\times T}$ at the base station and active sensors over the channel estimation phase of length $T$ are expressed as
\begin{align}
\mathbf{Y}&=\begin{bmatrix}\mathbf{y}[1]&\cdots&\mathbf{y}[T]\end{bmatrix}\notag\\
          &=\bar{\mathbf{G}}(\mathbf{S}\odot\bar{\mathbf{F}}\mathbf{X})+\bar{\mathbf{H}}\mathbf{X}+\mathbf{N}_{\mathrm{B}},\label{y}\\
\mathbf{Z}&=\begin{bmatrix}\mathbf{z}[1]&\cdots&\mathbf{z}[T]\end{bmatrix}\notag\\
          &=\bm{\Omega}\odot\mathrm{Q}(\bar{\mathbf{F}}\mathbf{X}+\mathbf{N}_{\mathrm{I}})\label{z}
\end{align}
using the notations $\mathbf{S}=[\mathbf{s}[1], \dots, \mathbf{s}[T]]$, $\bm{\Omega}=[\bm{\Omega}[1], \dots, \bm{\Omega}[T]]$, $\mathbf{X}=[\mathbf{x}[1], \dots, \mathbf{x}[T]]$, $\mathbf{N}_{\mathrm{B}}=[\mathbf{n}_{\mathrm{B}}[1], \dots, \mathbf{n}_{\mathrm{B}}[T]]$, and $\mathbf{N}_{\mathrm{I}}=[\mathbf{n}_{\mathrm{I}}[1], \dots, \mathbf{n}_{\mathrm{I}}[T]]$.

In addition, the semi-passive elements are configured as
\begin{equation}\label{norm}
\|\bm{\Omega}[t]\|_{0}=\begin{cases}N_{\mathrm{a}}&\text{for }t\in\mathcal{T}_{\mathrm{c}}\\0&\text{for }t\in\mathcal{T}_{\mathrm{d}}\end{cases},
\end{equation}
which means that $N_{\mathrm{a}}$ semi-passive elements are activated to the active sensing mode in the channel estimation phase and deactivated to the passive reflecting mode in the data transmission phase. The switching performance of switching networks that connect the IRS elements to RF chains is determined by the switching period $T_{\mathrm{sn}}\geq 1$ defined as the time required for the 0-1 pattern of $\bm{\Omega}[t]$ to change such that
\begin{equation}\label{support}
\bm{\Omega}[(i-1)T_{\mathrm{sn}}+1]=\cdots=\bm{\Omega}[iT_{\mathrm{sn}}]\text{ for }i\in\mathbb{N},
\end{equation}
and we define the switching frequency as $f_{\mathrm{sn}}=1/T_{\mathrm{sn}}\leq 1$. Therefore, the 0-1 pattern of $\bm{\Omega}[t]$ can change according to \eqref{norm} and \eqref{support}.

\subsection{Problem Formulation via Virtual Channel Representation}
We formulate the channel estimation problem using the virtual channel representation in conjunction with the channel model and signal model introduced in the previous subsections. To proceed, define the overcomplete dictionaries
\begin{align}
\mathbf{A}_{\mathrm{B}}&=\begin{bmatrix}\mathbf{a}_{\mathrm{B}}(\hat{\theta}_{1})&\cdots&\mathbf{a}_{\mathrm{B}}(\hat{\theta}_{M_{\mathrm{g}}})\end{bmatrix}\in\mathbb{C}^{M\times M_{\mathrm{g}}},\\
\mathbf{A}_{\mathrm{I}}&=\begin{bmatrix}\mathbf{a}_{\mathrm{I}}(\hat{\theta}_{1}^{\mathrm{A}}, \hat{\theta}_{1}^{\mathrm{Z}})&\cdots&\mathbf{a}_{\mathrm{I}}(\hat{\theta}_{N_{\mathrm{g}}}^{\mathrm{A}}, \hat{\theta}_{N_{\mathrm{g}}}^{\mathrm{Z}})\end{bmatrix}\in\mathbb{C}^{N\times N_{\mathrm{g}}}
\end{align}
over the predefined grids $\{\hat{\theta}_{m}\}_{m\in\llbracket M_{\mathrm{g}}\rrbracket}$ and $\{(\hat{\theta}_{n}^{\mathrm{A}}, \hat{\theta}_{n}^{\mathrm{Z}})\}_{n\in\llbracket N_{\mathrm{g}}\rrbracket}$ where $M_{\mathrm{g}}\geq M$ and $N_{\mathrm{g}}\geq N$ are the grid resolutions. Then, the virtual channel representation admits the transformations \cite{1033686}
\begin{align}
\bar{\mathbf{F}}&=\mathbf{A}_{\mathrm{I}}\mathbf{F},\\
\bar{\mathbf{G}}&=\mathbf{A}_{\mathrm{B}}\mathbf{G}\mathbf{A}_{\mathrm{I}}^{\mathrm{H}},\\
\bar{\mathbf{H}}&=\mathbf{A}_{\mathrm{B}}\mathbf{H}
\end{align}
where $\mathbf{F}\in\mathbb{C}^{N_{\mathrm{g}}\times K}$, $\mathbf{G}\in\mathbb{C}^{M_{\mathrm{g}}\times N_{\mathrm{g}}}$, and $\mathbf{H}\in\mathbb{C}^{M_{\mathrm{g}}\times K}$ are the equivalent angular-domain channels. In practice, the angular-domain channels are sparse because there are limited scatterers that constitute the channels in the mmWave band \cite{6847111, 7961152}.

Now, we can reexpress \eqref{y} and \eqref{z} using the virtual channel representation as
\begin{align}
\mathbf{Y}&=\mathbf{A}_{\mathrm{B}}\mathbf{G}\mathbf{A}_{\mathrm{I}}^{\mathrm{H}}(\mathbf{S}\odot\mathbf{A}_{\mathrm{I}}\mathbf{F}\mathbf{X})+\mathbf{A}_{\mathrm{B}}\mathbf{H}\mathbf{X}+\mathbf{N}_{\mathrm{B}},\label{yv}\\
\mathbf{Z}&=\bm{\Omega}\odot\mathrm{Q}(\mathbf{A}_{\mathrm{I}}\mathbf{F}\mathbf{X}+\mathbf{N}_{\mathrm{I}}),\label{zv}
\end{align}
and our goal is to estimate $\{\mathbf{F}, \mathbf{G}, \mathbf{H}\}$ from $\{\mathbf{Y}, \mathbf{Z}\}$. Since all the links are sparse, \eqref{yv} and \eqref{zv} is a combination of low-rank matrix factorization \cite{5197422} and low-rank matrix completion \cite{candes2009exact} problems, which are NP-hard in general. To the best of our knowledge, our work is the first attempt to jointly exploit the received signal at the base station and additional information acquired from the active sensors at the IRS to estimate the UE-IRS and IRS-BS links. In contrast, recent works \cite{9511813, 9529045} on channel estimation with semi-passive elements cannot be considered as semi-passive element-\textit{aided} but rather \textit{driven} because only the received signal at the semi-passive elements is leveraged. As a result, \cite{9511813, 9529045} require both uplink and downlink training signals to estimate the UE-IRS and BS-IRS links, whereas we only use uplink training signals to estimate all the links.

\section{Proposed VI-SBL-Based Channel Estimator}\label{section_3}
In this section, a VI-SBL-based channel estimator is proposed that performs approximate posterior inference on $\{\mathbf{F}, \mathbf{G}, \mathbf{H}\}$ from $\{\mathbf{Y}, \mathbf{Z}\}$ under the SBL framework \cite{10.1162/15324430152748236, 1315936}. In particular, we solve SBL via the variational free energy principle with the mean-field approximation \cite{4644060, MAL-001} to derive the posterior distributions of $\{\mathbf{F}, \mathbf{G}, \mathbf{H}\}$. Then, we can compute the posterior means from the approximate posterior distributions, or equivalently the MMSE estimates that minimize the channel estimation error.

\subsection{Pseudo-Measurement Model}
To facilitate the analysis, we propose a pseudo-measurement model
\begin{align}
\hat{\mathbf{Z}}&=\mathrm{Q}(\mathbf{U})\notag\\
                &=\mathrm{Q}(\bm{\Omega}\odot\mathbf{A}_{\mathrm{I}}\mathbf{F}\mathbf{X}+\mathbf{N}_{\mathrm{I}})\label{zh}
\end{align}
where the elements of $\hat{\mathbf{Z}}\in\mathbb{C}^{N\times T}$ corresponding to the nonzero pattern of $\bm{\Omega}$ are equal to the nonzero elements of $\mathbf{Z}$. In addition, the lower and upper pseudo-thresholds $\hat{\mathbf{Z}}^{\mathrm{lo}}\in\mathbb{C}^{N\times T}$ and $\hat{\mathbf{Z}}^{\mathrm{up}}\in\mathbb{C}^{N\times T}$ associated with $\hat{\mathbf{Z}}$ are defined as in \eqref{qu}. The elements of $\mathbf{Z}$ and $\hat{\mathbf{Z}}$ corresponding to the zeros of $\bm{\Omega}$, however, are not the same. Nevertheless, \eqref{zv} and \eqref{zh} are statistically equivalent because the elements of $\mathbf{Z}$ and $\hat{\mathbf{Z}}$ corresponding to the nonzero pattern of $\bm{\Omega}$ bear the same information about $\mathbf{F}$, while those corresponding to the zeros of $\bm{\Omega}$ have no meaningful information about $\mathbf{F}$ as evident from \eqref{zv} and \eqref{zh}. Therefore, the elements of $\hat{\mathbf{Z}}$, $\hat{\mathbf{Z}}^{\mathrm{lo}}$, and $\hat{\mathbf{Z}}^{\mathrm{up}}$ corresponding to the zeros of $\bm{\Omega}$---the elements that cannot be determined from $\mathbf{Z}$---can be assigned arbitrarily from the $2^{B}$ quantization intervals of the $B$-bit quantizer without loss of generality. Since the pseudo-measurement model is more convenient to deal with, we estimate $\{\mathbf{F}, \mathbf{G}, \mathbf{H}\}$ from $\{\mathbf{Y}, \hat{\mathbf{Z}}\}$ in \eqref{yv} and \eqref{zh} in the sequel.

\subsection{Hierarchical Bayesian Model of SBL Framework}
\begin{figure}[t]
\centering
\includegraphics[width=0.8\columnwidth]{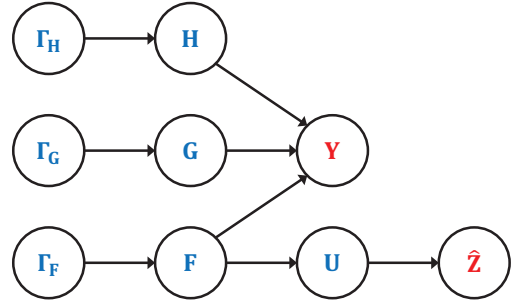}
\caption[caption]{The Bayesian network of a hierarchical Bayesian model where $\mathcal{Y}=\{\mathbf{Y}, \hat{\mathbf{Z}}\}$ is the measurement and $\mathcal{X}=\{\mathbf{F}, \mathbf{G}, \mathbf{H}, \bm{\Gamma}_{\mathbf{F}}, \bm{\Gamma}_{\mathbf{G}}, \bm{\Gamma}_{\mathbf{H}}, \mathbf{U}\}$ is the hidden variable. The arrows represent the conditional dependence between two random variables.}\label{figure_2}
\end{figure}

To account for the interaction among $\{\mathbf{F}, \mathbf{G}, \mathbf{H}, \mathbf{U}, \mathbf{Y}, \hat{\mathbf{Z}}\}$, we treat all the variables as random variables that constitute a hierarchical Bayesian model as shown in Fig. \ref{figure_2}. In addition, we introduce the random variables $\bm{\Gamma}_{\mathbf{F}}\in\mathbb{C}^{N_{\mathrm{g}}\times K}$, $\bm{\Gamma}_{\mathbf{G}}\in\mathbb{C}^{M_{\mathrm{g}}\times N_{\mathrm{g}}}$, and $\bm{\Gamma}_{\mathbf{H}}\in\mathbb{C}^{M_{\mathrm{g}}\times K}$ to capture the sparse nature of $\{\mathbf{F}, \mathbf{G}, \mathbf{H}\}$, which is the well-known SBL framework \cite{10.1162/15324430152748236, 1315936}. Before moving on, we introduce the equivalent vector forms of the measurement $\mathcal{Y}=\{\mathbf{y}, \hat{\mathbf{z}}\}$, hidden variable $\mathcal{X}=\{\mathbf{f}, \mathbf{g}, \mathbf{h}, \bm{\gamma}_{\mathbf{f}}, \bm{\gamma}_{\mathbf{g}}, \bm{\gamma}_{\mathbf{h}}, \mathbf{u}\}$, and pseudo-threshold $\{\hat{\mathbf{z}}^{\mathrm{lo}}, \hat{\mathbf{z}}^{\mathrm{up}}\}$ to facilitate the analysis. For example, $\mathbf{f}=\mathrm{vec}(\mathbf{F})$, $\hat{\mathbf{z}}^{\mathrm{lo}}=\mathrm{vec}(\hat{\mathbf{Z}}^{\mathrm{lo}})$, and $\hat{\mathbf{z}}^{\mathrm{up}}=\mathrm{vec}(\hat{\mathbf{Z}}^{\mathrm{up}})$.

In essence, the goal of SBL is to perform posterior inference on $\mathcal{X}$ from $\mathcal{Y}$ where the interaction among $\{\mathcal{X}, \mathcal{Y}\}$ is captured by the conditional distributions listed as follows. First, $\{\mathbf{f}, \mathbf{g}, \mathbf{h}\}$ conditioned on $\{\bm{\gamma}_{\mathbf{f}}, \bm{\gamma}_{\mathbf{g}}, \bm{\gamma}_{\mathbf{h}}\}$ are assumed to be Gaussian distributed as
\begin{align}
p(\mathbf{f}|\bm{\gamma}_{\mathbf{f}})&=\mathcal{CN}(\mathbf{f}|\mathbf{0}, \bm{\Gamma}_{\mathbf{f}}^{-1}),\label{pfr}\\
p(\mathbf{g}|\bm{\gamma}_{\mathbf{g}})&=\mathcal{CN}(\mathbf{g}|\mathbf{0}, \bm{\Gamma}_{\mathbf{g}}^{-1}),\label{pgr}\\
p(\mathbf{h}|\bm{\gamma}_{\mathbf{h}})&=\mathcal{CN}(\mathbf{h}|\mathbf{0}, \bm{\Gamma}_{\mathbf{h}}^{-1})\label{phr}
\end{align}
where $\bm{\Gamma}_{\mathbf{f}}=\mathrm{diag}(\bm{\gamma}_{\mathbf{f}})$, $\bm{\Gamma}_{\mathbf{g}}=\mathrm{diag}(\bm{\gamma}_{\mathbf{g}})$, and $\bm{\Gamma}_{\mathbf{h}}=\mathrm{diag}(\bm{\gamma}_{\mathbf{h}})$ are the precision matrices of the Gaussian distributions above. Meanwhile, the hyperpriors of the hyperparameters $\{\bm{\gamma}_{\mathbf{f}}, \bm{\gamma}_{\mathbf{g}}, \bm{\gamma}_{\mathbf{h}}\}$ are modeled as independent and identically distributed (i.i.d.) Gamma distributions
\begin{align}
p(\bm{\gamma}_{\mathbf{f}})&=\prod_{i}\mathrm{Gamma}(\gamma_{\mathbf{f}, i}|a, b)\notag\\
                           &=\prod_{i}\frac{b^{a}}{\Gamma(a)}\gamma_{\mathbf{f}, i}^{a-1}e^{-b\gamma_{\mathbf{f}, i}},\label{prf}\\
p(\bm{\gamma}_{\mathbf{g}})&=\prod_{i}\mathrm{Gamma}(\gamma_{\mathbf{g}, i}|a, b)\notag\\
                           &=\prod_{i}\frac{b^{a}}{\Gamma(a)}\gamma_{\mathbf{g}, i}^{a-1}e^{-b\gamma_{\mathbf{g}, i}},\\
p(\bm{\gamma}_{\mathbf{h}})&=\prod_{i}\mathrm{Gamma}(\gamma_{\mathbf{h}, i}|a, b)\notag\\
                           &=\prod_{i}\frac{b^{a}}{\Gamma(a)}\gamma_{\mathbf{h}, i}^{a-1}e^{-b\gamma_{\mathbf{h}, i}}
\end{align}
where $\Gamma(\cdot)$ is the Gamma function, and $a$ and $b$ are the shape and rate to be chosen. The reason for choosing the Gaussian-Gamma distribution is because $\{p(\bm{\gamma_{\mathbf{f}}}), p(\bm{\gamma_{\mathbf{g}}}), p(\bm{\gamma_{\mathbf{h}}})\}$ are the conjugate priors for the likelihood functions $\{p(\mathbf{f}|\bm{\gamma}_{\mathbf{f}}), p(\mathbf{g}|\bm{\gamma}_{\mathbf{g}}), p(\mathbf{h}|\bm{\gamma}_{\mathbf{h}})\}$ that make posterior inference tractable \cite{hoff2009first}. Furthermore, the Gaussian-Gamma distribution captures the sparse nature of $\{\mathbf{f}, \mathbf{g}, \mathbf{h}\}$ by making the marginal priors $\{p(\mathbf{f}), p(\mathbf{g}), p(\mathbf{h})\}$ Student's-$t$ distributions, which are sparsity-promoting priors under the appropriate choice of $a$ and $b$ \cite{10.1162/15324430152748236, 1315936}. In the simulation, we adhere to the convention that assumes uninformative priors by setting $a=b=10^{-6}$.

Moving on to the measurement model, the conditional distribution of $\mathbf{y}$ is
\begin{align}
&p(\mathbf{y}|\mathbf{f}, \mathbf{g}, \mathbf{h})=\notag\\
&\mathcal{CN}(\mathbf{y}|\mathrm{vec}(\mathbf{A}_{\mathrm{B}}\mathbf{G}\mathbf{A}_{\mathrm{I}}^{\mathrm{H}}(\mathbf{S}\odot\mathbf{A}_{\mathrm{I}}\mathbf{F}\mathbf{X})+\mathbf{A}_{\mathrm{B}}\mathbf{H}\mathbf{X}), \sigma_{\mathrm{B}}^{2}\mathbf{I}),\label{pyfgh}
\end{align}
which follows from \eqref{yv}. Likewise, the conditional distributions of $\mathbf{u}$ and $\hat{\mathbf{z}}$ in the pseudo-measurement model are
\begin{align}
      p(\mathbf{u}|\mathbf{f})=&\mathcal{CN}(\mathbf{u}|\mathrm{vec}(\bm{\Omega}\odot\mathbf{A}_{\mathrm{I}}\mathbf{F}\mathbf{X}), \sigma_{\mathrm{I}}^{2}\mathbf{I}),\label{puf}\\
p(\hat{\mathbf{z}}|\mathbf{u})=&\mathbb{I}(\hat{\mathbf{z}}=\mathrm{Q}(\mathbf{u}))\notag\\
                              =&\mathbb{I}(\mathrm{Re}(\hat{\mathbf{z}}^{\mathrm{lo}})\preceq\mathrm{Re}(\mathbf{u})\prec\mathrm{Re}(\hat{\mathbf{z}}^{\mathrm{up}}))\times\notag\\
                               &\mathbb{I}(\mathrm{Im}(\hat{\mathbf{z}}^{\mathrm{lo}})\preceq\mathrm{Im}(\mathbf{u})\prec\mathrm{Im}(\hat{\mathbf{z}}^{\mathrm{up}}))\label{pzu}
\end{align}
where \eqref{puf} and \eqref{pzu} come from \eqref{zh} and \eqref{qu}. The indicator function $\mathbb{I}(\cdot)$ is equal to one if the argument is true and zero otherwise.

Now, recall that the goal of SBL is to perform exact posterior inference on $\mathcal{X}$ from $\mathcal{Y}$ using \eqref{pfr}-\eqref{pzu}. The problem, however, is that computing $p(\mathcal{X}|\mathcal{Y})=p(\mathcal{X}, \mathcal{Y})/\int p(\mathcal{X}, \mathcal{Y})d\mathcal{X}$ is intractable in general. Therefore, we focus on finding the approximate posterior distribution using the VI approach.

\subsection{VI Approach to SBL}
First, we explain the idea behind VI with the mean-field approximation \cite{4644060, MAL-001}. Then, we derive the approximate posterior distributions, which enable us to compute the posterior means. To proceed, consider the decomposition of the log-evidence
\begin{align}
&\log p(\mathcal{Y})=\notag\\
&\underbrace{\int q(\mathcal{X})\log\frac{q(\mathcal{X})}{p(\mathcal{X}|\mathcal{Y})}d\mathcal{X}}_{=D_{\mathrm{KL}}(q\|p)}+\underbrace{\int q(\mathcal{X})\log\frac{p(\mathcal{X}, \mathcal{Y})}{q(\mathcal{X})}d\mathcal{X}}_{=\mathcal{L}(q)}\label{logpy}
\end{align}
where $q(\mathcal{X})$ is any probability distribution that approximates $p(\mathcal{X}|\mathcal{Y})$, $D_{\mathrm{KL}}(q\|p)$ is the Kullback–Leibler (KL) divergence that measures the distance between $q(\mathcal{X})$ and $p(\mathcal{X}|\mathcal{Y})$, and $\mathcal{L}(q)$ is known as the negative variational free energy. Then, we can minimize $D_{\mathrm{KL}}(q\|p)$ with respect to $q(\mathcal{X})$ by maximizing $\mathcal{L}(q)$ because $\log p(\mathcal{Y})$ is constant.

In general, the variational free energy minimization problem is intractable for a general class of $q(\mathcal{X})$. Therefore, we adopt the mean-field approximation by considering the class of $q(\mathcal{X})$ that assumes independence among the partition of $\mathcal{X}$ such that $q(\mathcal{X})=\prod_{i}q(\mathcal{X}_{i})$. Then, $\{q(\mathcal{X}_{i})\}_{\forall i}$ is the global minimum of the variational free energy minimization problem if and only if \cite{4644060, MAL-001}
\begin{equation}\label{qx}
q(\mathcal{X}_{i})=\frac{1}{Z_{i}}\exp\left\{\underbrace{\mathbb{E}_{\prod_{j\neq i}q(\mathcal{X}_{j})}\{\log p(\mathcal{Y}|\mathcal{X})p(\mathcal{X})\}}_{=\langle\log p(\mathcal{Y}|\mathcal{X})p(\mathcal{X})\rangle_{\mathcal{X}_{i}}}\right\}\text{ for }\forall i
\end{equation}
where $Z_{i}$ is the normalization constant that makes $q(\mathcal{X}_{i})$ a valid probability distribution. Here, $\langle\cdot\rangle_{\mathcal{X}_{i}}$ indicates the expectation with respect to $\prod_{j\neq i}q(\mathcal{X}_{j})$. In addition, we use the notation $\langle\cdot\rangle$ to indicate the expectation with respect to $q(\mathcal{X})$ in the sequel.

Then, our goal is to derive the functional forms of $\{q(\mathcal{X}_{i})\}_{\forall i}$ from \eqref{qx} that enable us to deduce the distributions of $\{q(\mathcal{X}_{i})\}_{\forall i}$ by inspection. Since the distributions of $\{q(\mathcal{X}_{i})\}_{\forall i}$ are coupled in an intertwined manner as evident from \eqref{qx}, however, $q(\mathcal{X}_{i})$ is updated for fixed $\{q(\mathcal{X}_{j})\}_{j\neq i}$ by cycling through $i$, which leads us to a local minimum. Now, we derive the functional forms of $\{q(\mathcal{X}_{i})\}_{\forall i}$ to determine the approximate posterior distributions for the partition $\mathcal{X}=\{\mathbf{f}, \mathbf{g}, \mathbf{h}, \bm{\gamma}_{\mathbf{f}}, \bm{\gamma}_{\mathbf{g}}, \bm{\gamma}_{\mathbf{h}}, \mathbf{u}\}$.

\textbf{1. Derivation of $q(\mathbf{u})$:} To derive $q(\mathbf{u})$, we only need to plug in \eqref{puf} and \eqref{pzu} to \eqref{qx}. The reason for not plugging in \eqref{pfr}-\eqref{pyfgh} to \eqref{qx} is because the expectations of \eqref{pfr}-\eqref{pyfgh} with respect to $\langle\cdot\rangle_{\mathbf{u}}$ are constants that do not affect the functional form of $q(\mathbf{u})$. In the posterior inference jargon, $\{\mathbf{f}, \hat{\mathbf{z}}\}$ is said to be the Markov blanket of $\mathbf{u}$ \cite{pearl1988probabilistic}. In the sequel, we identify the Markov blanket by inspection without further explanation.

To proceed, define
\begin{align}
\mathbf{b}_{\mathbf{uf}}&=\mathrm{vec}(\bm{\Omega}\odot\mathbf{A}_{\mathrm{I}}\mathbf{F}\mathbf{X})\notag\\
                        &=\mathrm{diag}(\mathrm{vec}(\bm{\Omega}))(\mathbf{X}^{\mathrm{T}}\otimes\mathbf{A}_{\mathrm{I}})\mathbf{f}\label{buf}
\end{align}
from \eqref{puf}. Then, plugging in \eqref{puf} and \eqref{pzu} to \eqref{qx} yields
\begin{align}
q(\mathbf{u})&\propto\exp\{\log p(\hat{\mathbf{z}}|\mathbf{u})+\langle\log p(\mathbf{u}|\mathbf{f})\rangle_{\mathbf{u}}\}\notag\\
             &\propto\mathbb{I}(\hat{\mathbf{z}}=\mathrm{Q}(\mathbf{u}))\exp\{\langle-\|\mathbf{u}-\mathbf{b}_{\mathbf{uf}}\|_{2}^{2}/\sigma_{\mathrm{I}}^{2}\rangle_{\mathbf{u}}\}\notag\\
             &\propto\mathbb{I}(\hat{\mathbf{z}}=\mathrm{Q}(\mathbf{u}))\exp\{-\|\mathbf{u}-\langle\mathbf{b}_{\mathbf{uf}}\rangle\|_{2}^{2}/\sigma_{\mathrm{I}}^{2}\}\notag\\
             &\propto\mathbb{I}(\hat{\mathbf{z}}=\mathrm{Q}(\mathbf{u}))\mathcal{CN}(\mathbf{u}|\langle\mathbf{b}_{\mathbf{uf}}\rangle, \sigma_{\mathrm{I}}^{2}\mathbf{I}),
\end{align}
or equivalently
\begin{align*}
&q(\mathrm{Re}(u_{i}))\propto\\
&\mathbb{I}(\mathrm{Re}(\hat{z}_{i}^{\mathrm{lo}})\leq\mathrm{Re}(u_{i})<\mathrm{Re}(\hat{z}_{i}^{\mathrm{up}}))\mathcal{N}\left(\mathrm{Re}(u_{i})|\mathrm{Re}(\langle b_{\mathbf{uf}, i}\rangle), \frac{\sigma_{\mathrm{I}}^{2}}{2}\right)
\end{align*}
for the real as well as the imaginary part, from which we recognize that $\mathbf{u}$ is truncated Gaussian distributed. The posterior mean of $\mathbf{u}$ has a well-known form \cite{7355388, 8310593, 9392374}
\begin{align}
&\langle\mathrm{Re}(u_{i})\rangle=\notag\\
&\quad\mathrm{Re}(\langle b_{\mathbf{uf}, i}\rangle)-\frac{\sigma_{\mathrm{I}}}{2}\times\frac{\phi(\mathrm{Re}(\beta_{i}))-\phi(\mathrm{Re}(\alpha_{i}))}{\Phi(\mathrm{Re}(\beta_{i}))-\Phi(\mathrm{Re}(\alpha_{i}))},\label{mur}\\
&\langle\mathrm{Im}(u_{i})\rangle=\notag\\
&\quad\mathrm{Im}(\langle b_{\mathbf{uf}, i}\rangle)-\frac{\sigma_{\mathrm{I}}}{2}\times\frac{\phi(\mathrm{Im}(\beta_{i}))-\phi(\mathrm{Im}(\alpha_{i}))}{\Phi(\mathrm{Im}(\beta_{i}))-\Phi(\mathrm{Im}(\alpha_{i}))}\label{mui}
\end{align}
where $\phi(\cdot)$ and $\Phi(\cdot)$ are the standard normal PDF and cumulative distribution function (CDF), $\alpha_{i}=(\hat{z}_{i}^{\mathrm{lo}}-\langle b_{\mathbf{uf}, i}\rangle)/(\sigma_{\mathrm{I}}/\sqrt{2})$, and $\beta_{i}=(\hat{z}_{i}^{\mathrm{up}}-\langle b_{\mathbf{uf}, i}\rangle)/(\sigma_{\mathrm{I}}/\sqrt{2})$. To compute the expressions above, we need $\langle\mathbf{b}_{\mathbf{uf}}\rangle$, which we can obtain from \eqref{buf} by replacing $\mathbf{f}$ with $\langle\mathbf{f}\rangle$.

\textbf{2. Derivation of $q(\mathbf{f})$:} First, define $\mathbf{A}_{\mathbf{fg}}$, $\mathbf{b}_{\mathbf{fh}}$, and $\mathbf{A}_{\mathbf{f}}$ from \eqref{pyfgh} and \eqref{puf} as \eqref{afg} and \eqref{af} at the bottom of the next page. Then, plugging in \eqref{pfr}, \eqref{pyfgh}, and \eqref{puf} to \eqref{qx} with some straightforward but tedious algebra leads to
\setcounter{equation}{33}
\begin{align}
q(\mathbf{f})\propto&\exp\{\langle\log p(\mathbf{y}|\mathbf{f}, \mathbf{g}, \mathbf{h})+\log p(\mathbf{u}|\mathbf{f})+\log p(\mathbf{f}|\bm{\gamma}_{\mathbf{f}})\rangle_{\mathbf{f}}\}\notag\\
             \propto&\exp\{\langle-\|\mathbf{y}-\mathbf{b}_{\mathbf{fh}}-\mathbf{A}_{\mathbf{fg}}\mathbf{f}\|_{2}^{2}/\sigma_{\mathrm{B}}^{2}\rangle_{\mathbf{f}}\}\times\notag\\
                    &\exp\{\langle-\|\mathbf{u}-\mathbf{A}_{\mathbf{f}}\mathbf{f}\|_{2}^{2}/\sigma_{\mathrm{I}}^{2}\rangle_{\mathbf{f}}\}\times\notag\\
                    &\exp\{\langle-\mathbf{f}^{\mathrm{H}}\bm{\Gamma}_{\mathbf{f}}\mathbf{f}\rangle_{\mathbf{f}}\}\notag\\
             \propto&\exp\{-(\mathbf{f}-\mathbf{m}_{\mathbf{f}})^{\mathrm{H}}\mathbf{C}_{\mathbf{f}}^{-1}(\mathbf{f}-\mathbf{m}_{\mathbf{f}})\},
\end{align}
which is a Gaussian distribution with
\begin{align}
\mathbf{m}_{\mathbf{f}}&=\mathbf{C}_{\mathbf{f}}\left(\frac{1}{\sigma_{\mathrm{B}}^{2}}\langle\mathbf{A}_{\mathbf{fg}}\rangle^{\mathrm{H}}(\mathbf{y}-\langle\mathbf{b}_{\mathbf{fh}}\rangle)+\frac{1}{\sigma_{\mathrm{I}}^{2}}\mathbf{A}_{\mathbf{f}}^{\mathrm{H}}\langle\mathbf{u}\rangle\right),\label{mf}\\
\mathbf{C}_{\mathbf{f}}&=\left(\frac{1}{\sigma_{\mathrm{B}}^{2}}\langle\mathbf{A}_{\mathbf{fg}}^{\mathrm{H}}\mathbf{A}_{\mathbf{fg}}\rangle+\frac{1}{\sigma_{\mathrm{I}}^{2}}\mathbf{A}_{\mathbf{f}}^{\mathrm{H}}\mathbf{A}_{\mathbf{f}}+\langle\bm{\Gamma}_{\mathbf{f}}\rangle\right)^{-1}.\label{cf}
\end{align}
The posterior mean and covariance above require $\langle\mathbf{A}_{\mathbf{fg}}\rangle$, $\langle\mathbf{b}_{\mathbf{fh}}\rangle$, and $\langle\mathbf{A}_{\mathbf{fg}}^{\mathrm{H}}\mathbf{A}_{\mathbf{fg}}\rangle$ to be computed, from which the first two can be obtained from \eqref{afg} by substituting $\mathbf{G}$ and $\mathbf{h}$ with $\langle\mathbf{G}\rangle$ and $\langle\mathbf{h}\rangle$. The expression for $\langle\mathbf{A}_{\mathbf{fg}}^{\mathrm{H}}\mathbf{A}_{\mathbf{fg}}\rangle$ is provided in Appendix \ref{appendix_a} as the derivation is more involved.

\begin{figure*}[b]
\hrulefill
\begin{align}
\setcounter{equation}{31}
&\mathrm{vec}(\mathbf{A}_{\mathrm{B}}\mathbf{G}\mathbf{A}_{\mathrm{I}}^{\mathrm{H}}(\mathbf{S}\odot\mathbf{A}_{\mathrm{I}}\mathbf{F}\mathbf{X})+\mathbf{A}_{\mathrm{B}}\mathbf{H}\mathbf{X})=\underbrace{(\mathbf{I}_{T}\otimes\mathbf{A}_{\mathrm{B}}\mathbf{G}\mathbf{A}_{\mathrm{I}}^{\mathrm{H}})\mathrm{diag}(\mathrm{vec}(\mathbf{S}))(\mathbf{X}^{\mathrm{T}}\otimes\mathbf{A}_{\mathrm{I}})}_{=\mathbf{A}_{\mathbf{fg}}}\mathbf{f}+\underbrace{(\mathbf{X}^{\mathrm{T}}\otimes\mathbf{A}_{\mathrm{B}})\mathbf{h}}_{=\mathbf{b}_{\mathbf{fh}}},\label{afg}\\
&\mathrm{vec}(\bm{\Omega}\odot\mathbf{A}_{\mathrm{I}}\mathbf{F}\mathbf{X})=\underbrace{\mathrm{diag}(\mathrm{vec}(\bm{\Omega}))(\mathbf{X}^{\mathrm{T}}\otimes\mathbf{A}_{\mathrm{I}})}_{=\mathbf{A}_{\mathbf{f}}}\mathbf{f},\label{af}\\
\setcounter{equation}{36}
&\mathrm{vec}(\mathbf{A}_{\mathrm{B}}\mathbf{G}\mathbf{A}_{\mathrm{I}}^{\mathrm{H}}(\mathbf{S}\odot\mathbf{A}_{\mathrm{I}}\mathbf{F}\mathbf{X})+\mathbf{A}_{\mathrm{B}}\mathbf{H}\mathbf{X})=\underbrace{((\mathbf{S}^{\mathrm{T}}\odot\mathbf{X}^{\mathrm{T}}\mathbf{F}^{\mathrm{T}}\mathbf{A}_{\mathrm{I}}^{\mathrm{T}})\mathbf{A}_{\mathrm{I}}^{*}\otimes\mathbf{A}_{\mathrm{B}})}_{=\mathbf{A}_{\mathbf{gf}}}\mathbf{g}+\underbrace{(\mathbf{X}^{\mathrm{T}}\otimes\mathbf{A}_{\mathrm{B}})\mathbf{h}}_{=\mathbf{b}_{\mathbf{gh}}},\label{agf}\\
\setcounter{equation}{40}
&\mathrm{vec}(\mathbf{A}_{\mathrm{B}}\mathbf{G}\mathbf{A}_{\mathrm{I}}^{\mathrm{H}}(\mathbf{S}\odot\mathbf{A}_{\mathrm{I}}\mathbf{F}\mathbf{X})+\mathbf{A}_{\mathrm{B}}\mathbf{H}\mathbf{X})=\underbrace{(\mathbf{X}^{\mathrm{T}}\otimes\mathbf{A}_{\mathrm{B}})}_{=\mathbf{A}_{\mathbf{h}}}\mathbf{h}+\underbrace{((\mathbf{S}^{\mathrm{T}}\odot\mathbf{X}^{\mathrm{T}}\mathbf{F}^{\mathrm{T}}\mathbf{A}_{\mathrm{I}}^{\mathrm{T}})\mathbf{A}_{\mathrm{I}}^{*}\otimes\mathbf{A}_{\mathrm{B}})\mathbf{g}}_{=\mathbf{b}_{\mathbf{hfg}}}\label{ah}
\end{align}
\end{figure*}

\textbf{3. Derivation of $q(\mathbf{g})$:} First, we define $\mathbf{A}_{\mathbf{gf}}$ and $\mathbf{b}_{\mathbf{gh}}$ from \eqref{pyfgh} as \eqref{agf} at the bottom of the next page. Then, expanding \eqref{qx} after plugging in \eqref{pgr} and \eqref{pyfgh} gives
\setcounter{equation}{37}
\begin{align}
q(\mathbf{g})\propto&\exp\{\langle\log p(\mathbf{y}|\mathbf{f}, \mathbf{g}, \mathbf{h})+\log p(\mathbf{g}|\bm{\gamma}_{\mathbf{g}})\rangle_{\mathbf{g}}\}\notag\\
             \propto&\exp\{\langle-\|\mathbf{y}-\mathbf{b}_{\mathbf{gh}}-\mathbf{A}_{\mathbf{gf}}\mathbf{g}\|_{2}^{2}/\sigma_{\mathrm{B}}^{2}\rangle_{\mathbf{g}}\}\times\notag\\
                    &\exp\{\langle-\mathbf{g}^{\mathrm{H}}\bm{\Gamma}_{\mathbf{g}}\mathbf{g}\rangle_{\mathbf{g}}\}\notag\\
             \propto&\exp\{-(\mathbf{g}-\mathbf{m}_{\mathbf{g}})^{\mathrm{H}}\mathbf{C}_{\mathbf{g}}^{-1}(\mathbf{g}-\mathbf{m}_{\mathbf{g}})\},
\end{align}
which is recognized as a Gaussian distribution associated with the posterior mean and covariance
\begin{align}
\mathbf{m}_{\mathbf{g}}&=\mathbf{C}_{\mathbf{g}}\left(\frac{1}{\sigma_{\mathrm{B}}^{2}}\langle\mathbf{A}_{\mathbf{gf}}\rangle^{\mathrm{H}}(\mathbf{y}-\langle\mathbf{b}_{\mathbf{gh}}\rangle)\right),\label{mg}\\
\mathbf{C}_{\mathbf{g}}&=\left(\frac{1}{\sigma_{\mathrm{B}}^{2}}\langle\mathbf{A}_{\mathbf{gf}}^{\mathrm{H}}\mathbf{A}_{\mathbf{gf}}\rangle+\langle\bm{\Gamma}_{\mathbf{g}}\rangle\right)^{-1}.\label{cg}
\end{align}
Again, computing the expressions above requires $\langle\mathbf{A}_{\mathbf{gf}}\rangle$ and $\langle\mathbf{b}_{\mathbf{gh}}\rangle$ as well as $\langle\mathbf{A}_{\mathbf{gf}}^{\mathrm{H}}\mathbf{A}_{\mathbf{gf}}\rangle$, from which the terms corresponding to the first moment are given by \eqref{agf} after replacing $\mathbf{F}$ and $\mathbf{h}$ with $\langle\mathbf{F}\rangle$ and $\langle\mathbf{h}\rangle$. The explicit form of $\langle\mathbf{A}_{\mathbf{gf}}^{\mathrm{H}}\mathbf{A}_{\mathbf{gf}}\rangle$ with a detailed derivation is provided in Appendix \ref{appendix_b}.

\textbf{4. Derivation of $q(\mathbf{h})$:} First, define $\mathbf{A}_{\mathbf{h}}$ and $\mathbf{b}_{\mathbf{hfg}}$ from \eqref{pyfgh} as \eqref{ah} at the bottom of the next page. By plugging in \eqref{phr} and \eqref{pyfgh} to \eqref{qx}, we arrive at
\setcounter{equation}{41}
\begin{align}
q(\mathbf{h})\propto&\exp\{\langle\log p(\mathbf{y}|\mathbf{f}, \mathbf{g}, \mathbf{h})+\log p(\mathbf{h}|\bm{\gamma}_{\mathbf{h}})\rangle_{\mathbf{h}}\}\notag\\
             \propto&\exp\{\langle-\|\mathbf{y}-\mathbf{b}_{\mathbf{hfg}}-\mathbf{A}_{\mathbf{h}}\mathbf{h}\|_{2}^{2}/\sigma_{\mathrm{B}}^{2}\rangle_{\mathbf{h}}\}\times\notag\\
                    &\exp\{\langle-\mathbf{h}^{\mathrm{H}}\bm{\Gamma}_{\mathbf{h}}\mathbf{h}\rangle_{\mathbf{h}}\}\notag\\
             \propto&\exp\{-(\mathbf{h}-\mathbf{m}_{\mathbf{h}})^{\mathrm{H}}\mathbf{C}_{\mathbf{h}}^{-1}(\mathbf{h}-\mathbf{m}_{\mathbf{h}})\},
\end{align}
which implies that $\mathbf{h}$ is Gaussian distributed parameterized by
\begin{align}
\mathbf{m}_{\mathbf{h}}&=\mathbf{C}_{\mathbf{h}}\left(\frac{1}{\sigma_{\mathrm{B}}^{2}}\mathbf{A}_{\mathbf{h}}^{\mathrm{H}}(\mathbf{y}-\langle\mathbf{b}_{\mathbf{hfg}}\rangle)\right),\label{mh}\\
\mathbf{C}_{\mathbf{h}}&=\left(\frac{1}{\sigma_{\mathrm{B}}^{2}}\mathbf{A}_{\mathbf{h}}^{\mathrm{H}}\mathbf{A}_{\mathbf{h}}+\langle\bm{\Gamma}_{\mathbf{h}}\rangle\right)^{-1}.\label{ch}
\end{align}
The posterior mean and covariance above require $\langle\mathbf{b}_{\mathbf{hfg}}\rangle$ to be computed, which can be obtained from \eqref{ah} by substituting $\mathbf{F}$ and $\mathbf{g}$ with $\langle\mathbf{F}\rangle$ and $\langle\mathbf{g}\rangle$.

\textbf{5. Derivation of $q(\bm{\gamma}_{\mathbf{f}})$:} Let us plug in \eqref{pfr} and \eqref{prf} to \eqref{qx} to obtain
\begin{align}
q(\bm{\gamma}_{\mathbf{f}})\propto&\exp\{\langle\log p(\mathbf{f}|\bm{\gamma}_{\mathbf{f}})\rangle_{\bm{\gamma}_{\mathbf{f}}}+\log p(\bm{\gamma}_{\mathbf{f}})\}\notag\\
                           \propto&\exp\{\log\det(\bm{\Gamma}_{\mathbf{f}})-\langle\mathbf{f}^{\mathrm{H}}\bm{\Gamma}_{\mathbf{f}}\mathbf{f}\rangle_{\bm{\gamma}_{\mathbf{f}}}+\log p(\bm{\gamma}_{\mathbf{f}})\}\notag\\
                           \propto&\prod_{i}\gamma_{\mathbf{f}, i}\times\exp\{-\langle|f_{i}|^{2}\rangle\gamma_{\mathbf{f}, i}\}\times\gamma_{\mathbf{f}, i}^{a-1}\exp\{-b\gamma_{\mathbf{f}, i}\}\notag\\
                           \propto&\prod_{i}\gamma_{\mathbf{f}, i}^{(a+1)-1}\exp\{-(b+\langle|f_{i}|^{2}\rangle)\gamma_{\mathbf{f}, i}\},
\end{align}
which is recognized as a product of Gamma distributions associated with the same posterior shape $\bar{a}=a+1$ and different rates $\bar{b}_{\mathbf{f}, i}=b+\langle|f_{i}|^{2}\rangle=b+[\mathbf{C}_{\mathbf{f}}+\mathbf{m}_{\mathbf{f}}\mathbf{m}_{\mathbf{f}}^{\mathrm{H}}]_{i, i}$. Therefore, the posterior mean of $\bm{\gamma}_{\mathbf{f}}$ is
\begin{equation}\label{mrf}
\langle\gamma_{\mathbf{f}, i}\rangle=\frac{\bar{a}}{\bar{b}_{\mathbf{f}, i}}.
\end{equation}

\textbf{6. Derivation of $q(\bm{\gamma}_{\mathbf{g}})$:} Since the priors $\{p(\mathbf{f}|\bm{\gamma}_{\mathbf{f}}), p(\mathbf{g}|\bm{\gamma}_{\mathbf{g}})\}$ as well as the hyperpriors $\{p(\bm{\gamma}_{\mathbf{f}}), p(\bm{\gamma}_{\mathbf{g}})\}$ are i.i.d., the derivation of $q(\bm{\gamma}_{\mathbf{g}})$ leads to the same functional form, which is a product of Gamma distributions with the same posterior shape $\bar{a}=a+1$ and different rates $\bar{b}_{\mathbf{g}, i}=b+\langle|g_{i}|^{2}\rangle=b+[\mathbf{C}_{\mathbf{g}}+\mathbf{m}_{\mathbf{g}}\mathbf{m}_{\mathbf{g}}^{\mathrm{H}}]_{i, i}$. Therefore, the posterior mean of $\bm{\gamma}_{\mathbf{g}}$ is
\begin{equation}\label{mrg}
\langle\gamma_{\mathbf{g}, i}\rangle=\frac{\bar{a}}{\bar{b}_{\mathbf{g}, i}}.
\end{equation}

\textbf{7. Derivation of $q(\bm{\gamma}_{\mathbf{h}})$:} By the same argument as in the derivation of $q(\bm{\gamma}_{\mathbf{g}})$, we conclude that $q(\bm{\gamma}_{\mathbf{h}})$ is a product of Gamma distributions parameterized by the same posterior shape $\bar{a}=a+1$ and different rates $\bar{b}_{\mathbf{h}, i}=b+\langle|h_{i}|^{2}\rangle=b+[\mathbf{C}_{\mathbf{h}}+\mathbf{m}_{\mathbf{h}}\mathbf{m}_{\mathbf{h}}^{\mathrm{H}}]_{i, i}$, from which the posterior mean of $\bm{\gamma}_{\mathbf{h}}$ is deduced as
\begin{equation}\label{mrh}
\langle\gamma_{\mathbf{h}, i}\rangle=\frac{\bar{a}}{\bar{b}_{\mathbf{h}, i}}.
\end{equation}

The VI-SBL-based channel estimator developed until now is summarized in Algorithm \ref{algorithm_1}. In essence, Algorithm \ref{algorithm_1} performs approximate posterior inference by updating $\{q(\mathbf{f}), q(\mathbf{g}), q(\mathbf{h}), q(\bm{\gamma}_{\mathbf{f}}), q(\bm{\gamma}_{\mathbf{g}}), q(\bm{\gamma}_{\mathbf{h}}), q(\mathbf{u})\}$ iteratively. The MMSE estimates of all the links are given by
\begin{align}
\hat{\bar{\mathbf{F}}}&=\mathbf{A}_{\mathrm{I}}\mathrm{reshape}(\mathbf{m}_{\mathbf{f}}, [N_{\mathrm{g}}, K]),\\
\hat{\bar{\mathbf{G}}}&=\mathbf{A}_{\mathrm{B}}\mathrm{reshape}(\mathbf{m}_{\mathbf{g}}, [M_{\mathrm{g}}, N_{\mathrm{g}}])\mathbf{A}_{\mathrm{I}}^{\mathrm{H}},\\
\hat{\bar{\mathbf{H}}}&=\mathbf{A}_{\mathrm{B}}\mathrm{reshape}(\mathbf{m}_{\mathbf{h}}, [M_{\mathrm{g}}, K])
\end{align}
where $\mathrm{reshape}(\mathbf{a}, [m, n])$ reshapes $\mathbf{a}$ to a matrix of size $m\times n$ that preserves the columnwise ordering. To investigate the convergence of Algorithm \ref{algorithm_1}, recall that Algorithm \ref{algorithm_1} tackles \eqref{qx} by solving $q(\mathcal{X}_{i})$ for fixed $\{q(\mathcal{X}_{j})\}_{j\neq i}$ in a cycling manner instead of jointly solving for $\{q(\mathcal{X}_{i})\}_{\forall i}$. Since \eqref{qx} defines the global minimum of the variational free energy minimization problem \cite{4644060, MAL-001}, each line in Algorithm \ref{algorithm_1} at least reduces the variational free energy. By noting that the variational free energy is lower-bounded by the negative log-evidence as evident from \eqref{logpy}, we conclude that Algorithm \ref{algorithm_1} converges to a local minimum of the variational free energy problem.

\begin{algorithm}[t]
\caption{VI-SBL-based channel estimator}\label{algorithm_1}
\begin{algorithmic}[1]
\Require $\mathbf{y}$, $\hat{\mathbf{z}}$
\Ensure $\mathbf{m}_{\mathbf{f}}$, $\mathbf{m}_{\mathbf{g}}$, $\mathbf{m}_{\mathbf{h}}$, or equivalently $\langle\mathbf{f}\rangle$, $\langle\mathbf{g}\rangle$, $\langle\mathbf{h}\rangle$
\State Set the parameters $a$ and $b$ for the hyperpriors
\State Initialize $\mathbf{m}_{\mathbf{f}}$, $\mathbf{C}_{\mathbf{f}}$, $\mathbf{m}_{\mathbf{h}}$, $\mathbf{C}_{\mathbf{h}}$, $\langle\bm{\gamma}_{\mathbf{f}}\rangle$, $\langle\bm{\gamma}_{\mathbf{g}}\rangle$, $\langle\bm{\gamma}_{\mathbf{h}}\rangle$, and $\langle\mathbf{u}\rangle$
\While {termination condition}
\State Update $\mathbf{m}_{\mathbf{g}}$ and $\mathbf{C}_{\mathbf{g}}$ according to \eqref{mg} and \eqref{cg}
\State Update $\langle\bm{\gamma}_{\mathbf{g}}\rangle$ according to \eqref{mrg}
\State Update $\mathbf{m}_{\mathbf{h}}$ and $\mathbf{C}_{\mathbf{h}}$ according to \eqref{mh} and \eqref{ch}
\State Update $\langle\bm{\gamma}_{\mathbf{h}}\rangle$ according to \eqref{mrh}
\State Update $\mathbf{m}_{\mathbf{f}}$ and $\mathbf{C}_{\mathbf{f}}$ according to \eqref{mf} and \eqref{cf}
\State Update $\langle\bm{\gamma}_{\mathbf{f}}\rangle$ according to \eqref{mrf}
\State Update $\langle\mathbf{u}\rangle$ according to \eqref{mur} and \eqref{mui}
\EndWhile
\end{algorithmic}
\end{algorithm}

The problem, however, is that poor initialization can result in a bad local minimum. The simulation results showed that a good initialization strategy is to perform ``initial" posterior inference on a subset of the parameters in advance, and then use the coarse estimates obtained in this stage to initialize the parameters. The initial posterior inference we consider is the UE-RIS link estimation and UE-BS link estimation. In particular, the UE-IRS link initialization proceeds by performing initial posterior inference on $\{\mathbf{f}, \bm{\gamma}_{\mathbf{f}}, \mathbf{u}\}$ based only on the measurement $\hat{\mathbf{z}}$ to initialize $\{\mathbf{m}_{\mathbf{f}}, \mathbf{C}_{\mathbf{f}}, \langle\bm{\gamma}_{\mathbf{f}}\rangle, \langle\mathbf{u}\rangle\}$ via Lines 8-10. For the UE-BS link initialization, we first turn off the passive reflecting elements for a short duration during the channel estimation phase as shown in \eqref{v}. Then, the UE-BS link initialization proceeds by performing initial posterior inference on $\{\mathbf{h}, \bm{\gamma}_{\mathbf{h}}\}$ based on a subset of $\mathbf{y}$ corresponding to the time slots the passive reflecting elements were turned off as in \eqref{v} to initialize $\{\mathbf{m}_{\mathbf{h}}, \mathbf{C}_{\mathbf{h}}, \langle\bm{\gamma}_{\mathbf{h}}\rangle\}$ via Lines 6-7. Since each initial posterior inference is equivalent to the conventional direct link channel estimation, i.e., UE-IRS link and UE-BS link, the initial posterior inference can be performed without much difficulty (even though the estimates obtained in this stage are very coarse due to the extremely small size of the measurements used, thus using them only for initialization and preventing us from using them as final estimates). The remaining parameter $\langle\bm{\gamma}_{\mathbf{g}}\rangle$, which is not associated with neither the UE-IRS nor UE-BS link initialization, can be initialized as its prior mean $\mathbb{E}\{\bm{\gamma}_{\mathbf{g}}\}=a/b\times\mathbf{1}$. By adopting the proposed initialization strategy, Algorithm \ref{algorithm_1} was demonstrated to produce good results.

\subsection{Complexity Reduction via Mean-Field Approximation}
The complexity of VI-SBL is mainly attributed to the matrix inversions in \eqref{cf}, \eqref{cg}, and \eqref{ch} where the matrices to be inverted are of sizes $N_{\mathrm{g}}K\times N_{\mathrm{g}}K$, $M_{\mathrm{g}}N_{\mathrm{g}}\times M_{\mathrm{g}}N_{\mathrm{g}}$, and $M_{\mathrm{g}}K\times M_{\mathrm{g}}K$. Therefore, the complexity of VI-SBL is $\mathcal{O}(N_{\mathrm{g}}^{3}K^{3}+M_{\mathrm{g}}^{3}N_{\mathrm{g}}^{3}+M_{\mathrm{g}}^{3}K^{3})$. To reduce the complexity of VI-SBL, a mean-field approximation-based solution is proposed that converts a large matrix inversion to many small matrix inversions.

In particular, we consider the mean-field approximation that splits $\{\mathbf{f}, \bm{\gamma}_{\mathbf{f}}\}$, $\{\mathbf{g}, \bm{\gamma}_{\mathbf{g}}\}$, and $\{\mathbf{h}, \bm{\gamma}_{\mathbf{h}}\}$ to $S_{\mathbf{f}}$, $S_{\mathbf{g}}$, and $S_{\mathbf{h}}$ subvectors. For example, the partition of $\{\mathbf{f}, \bm{\gamma}_{\mathbf{f}}\}$ can be written as $\mathbf{f}^{\mathrm{T}}=[\mathbf{f}_{[1]}^{\mathrm{T}}, \dots, \mathbf{f}_{[S_{\mathbf{f}}]}^{\mathrm{T}}]$ and $\bm{\gamma}_{\mathbf{f}}^{\mathrm{T}}=[\bm{\gamma}_{\mathbf{f}[1]}^{\mathrm{T}}, \dots, \bm{\gamma}_{\mathbf{f}[S_{\mathbf{f}}]}^{\mathrm{T}}]$ where $\mathbf{f}_{[i]}\in\mathbb{C}^{N_{\mathrm{g}}K/S_{\mathbf{f}}}$ and $\bm{\gamma}_{\mathbf{f}[i]}\in\mathbb{C}^{N_{\mathrm{g}}K/S_{\mathbf{f}}}$, and the same logic holds for $\{\mathbf{g}, \bm{\gamma}_{\mathbf{g}}\}$ and $\{\mathbf{h}, \bm{\gamma}_{\mathbf{h}}\}$. Then, VI-SBL performs posterior inference on the partition
\begin{align}
&q(\mathcal{X})=\notag\\
&q(\mathbf{u})\prod_{i=1}^{S_{\mathbf{f}}}q(\mathbf{f}_{[i]})q(\bm{\gamma}_{\mathbf{f}[i]})\prod_{j=1}^{S_{\mathbf{g}}}q(\mathbf{g}_{[j]})q(\bm{\gamma}_{\mathbf{g}[j]})\prod_{k=1}^{S_{\mathbf{h}}}q(\mathbf{h}_{[k]})q(\bm{\gamma}_{\mathbf{h}[k]})
\end{align}
from $\{\mathbf{y}, \hat{\mathbf{z}}\}$.

Now, we present how the update rules in \eqref{mf}, \eqref{cf}, \eqref{mg}, \eqref{cg}, \eqref{mh}, and \eqref{ch} are modified after the proposed mean-field approximation is applied. As a preliminary, let us introduce some shorthand notations. First, $\mathbf{f}_{[i]}^{\mathrm{c}}$ is defined as the vector obtained by removing $\mathbf{f}_{[i]}$ from $\mathbf{f}$. Likewise, $\mathbf{A}_{\mathbf{f}[i]\mathbf{g}}$ and $\mathbf{A}_{\mathbf{f}[i]\mathbf{g}}^{\mathrm{c}}$ are defined as the matrices obtained by retaining the columns of $\mathbf{A}_{\mathbf{fg}}$ in \eqref{afg} corresponding to $\mathbf{f}_{[i]}$ and $\mathbf{f}_{[i]}^{\mathrm{c}}$. The same logic can be extended to denote any vectors and matrices obtained by pruning. As an exception, the pruned versions of the precision matrices are denoted by $\bm{\Gamma}_{\mathbf{f}[i]}=\mathrm{diag}(\bm{\gamma}_{\mathbf{f}[i]})$, $\bm{\Gamma}_{\mathbf{g}[i]}=\mathrm{diag}(\bm{\gamma}_{\mathbf{g}[i]})$, and $\bm{\Gamma}_{\mathbf{h}[i]}=\mathrm{diag}(\bm{\gamma}_{\mathbf{h}[i]})$.

Then, the new update rules for \eqref{mf}, \eqref{cf}, \eqref{mg}, \eqref{cg}, \eqref{mh}, and \eqref{ch} are given by \eqref{mfi}-\eqref{mhi} at the bottom of the page. Since \eqref{mfi} repeats an $N_{\mathrm{g}}K/S_{\mathbf{f}}\times N_{\mathrm{g}}K/S_{\mathbf{f}}$ matrix inversion $S_{\mathbf{f}}$ times, the overall complexity is $\mathcal{O}(N_{\mathrm{g}}^{3}K^{3}/S_{\mathbf{f}}^{2})$. The same logic applies to \eqref{mgi} and \eqref{mhi}. As a result, the overall complexity of VI-SBL reduces to $\mathcal{O}(N_{\mathrm{g}}^{3}K^{3}/S_{\mathbf{f}}^{2}+M_{\mathrm{g}}^{3}N_{\mathrm{g}}^{3}/S_{\mathbf{g}}^{2}+M_{\mathrm{g}}^{3}K^{3}/S_{\mathbf{h}}^{2})$.

\begin{figure*}[b]
\hrulefill
\setcounter{equation}{52}
\begin{align}
\mathbf{m}_{\mathbf{f}[i]}&=\mathbf{C}_{\mathbf{f}[i]}\bigg(\frac{1}{\sigma_{\mathrm{B}}^{2}}\bigg\{\langle\mathbf{A}_{\mathbf{f}[i]\mathbf{g}}\rangle^{\mathrm{H}}(\mathbf{y}-\langle\mathbf{b}_{\mathbf{fh}}\rangle)-\langle\mathbf{A}_{\mathbf{f}[i]\mathbf{g}}^{\mathrm{H}}\mathbf{A}_{\mathbf{f}[i]\mathbf{g}}^{\mathrm{c}}\rangle\langle\mathbf{f}_{[i]}^{\mathrm{c}}\rangle\bigg\}+\frac{1}{\sigma_{\mathrm{I}}^{2}}\mathbf{A}_{\mathbf{f}[i]}^{\mathrm{H}}(\langle\mathbf{u}\rangle-\mathbf{A}_{\mathbf{f}[i]}^{\mathrm{c}}\langle\mathbf{f}_{[i]}^{\mathrm{c}}\rangle)\bigg),\notag\\
\mathbf{C}_{\mathbf{f}[i]}&=\left(\frac{1}{\sigma_{\mathrm{B}}^{2}}\langle\mathbf{A}_{\mathbf{f}[i]\mathbf{g}}^{\mathrm{H}}\mathbf{A}_{\mathbf{f}[i]\mathbf{g}}\rangle+\frac{1}{\sigma_{\mathrm{I}}^{2}}\mathbf{A}_{\mathbf{f}[i]}^{\mathrm{H}}\mathbf{A}_{\mathbf{f}[i]}+\langle\bm{\Gamma}_{\mathbf{f}[i]}\rangle\right)^{-1}\text{ for }i\in\llbracket S_{\mathbf{f}}\rrbracket,\label{mfi}\\
\mathbf{m}_{\mathbf{g}[i]}&=\mathbf{C}_{\mathbf{g}[i]}\bigg(\frac{1}{\sigma_{\mathrm{B}}^{2}}\bigg\{\langle\mathbf{A}_{\mathbf{g}[i]\mathbf{f}}\rangle^{\mathrm{H}}(\mathbf{y}-\langle\mathbf{b}_{\mathbf{gh}}\rangle)-\langle\mathbf{A}_{\mathbf{g}[i]\mathbf{f}}^{\mathrm{H}}\mathbf{A}_{\mathbf{g}[i]\mathbf{f}}^{\mathrm{c}}\rangle\langle\mathbf{g}_{[i]}^{\mathrm{c}}\rangle\bigg\}\bigg),\notag\\
\mathbf{C}_{\mathbf{g}[i]}&=\left(\frac{1}{\sigma_{\mathrm{B}}^{2}}\langle\mathbf{A}_{\mathbf{g}[i]\mathbf{f}}^{\mathrm{H}}\mathbf{A}_{\mathbf{g}[i]\mathbf{f}}\rangle+\langle\bm{\Gamma}_{\mathbf{g}[i]}\rangle\right)^{-1}\text{ for }i\in\llbracket S_{\mathbf{g}}\rrbracket,\label{mgi}\\
\mathbf{m}_{\mathbf{h}[i]}&=\mathbf{C}_{\mathbf{h}[i]}\left(\frac{1}{\sigma_{\mathrm{B}}^{2}}\mathbf{A}_{\mathbf{h}[i]}^{\mathrm{H}}(\mathbf{y}-\langle\mathbf{b}_{\mathbf{hfg}}\rangle-\mathbf{A}_{\mathbf{h}[i]}^{\mathrm{c}}\langle\mathbf{h}_{[i]}^{\mathrm{c}}\rangle)\right),\quad \mathbf{C}_{\mathbf{h}[i]}=\left(\frac{1}{\sigma_{\mathrm{B}}^{2}}\mathbf{A}_{\mathbf{h}[i]}^{\mathrm{H}}\mathbf{A}_{\mathbf{h}[i]}+\langle\bm{\Gamma}_{\mathbf{h}[i]}\rangle\right)^{-1}\text{ for }i\in\llbracket S_{\mathbf{h}}\rrbracket\label{mhi}
\end{align}
\end{figure*}

\section{Simulation Results}\label{section_4}
In this section, we evaluate the performance of the proposed VI-SBL-based channel estimator based on the channel estimation error and energy efficiency. The baselines are state-of-the-art compressed sensing-based channel estimators for IRS-aided mmWave massive MIMO systems with passive reflecting elements. To implement the baselines, the UE-IRS-BS link estimation problem is reformulated as a sparse recovery problem as proposed in \cite{9103231}. Then, the problem is solved using the generalized approximate MP (GAMP) \cite{6033942}, vector approximate MP (VAMP) \cite{7869633}, SBL \cite{10.1162/15324430152748236}, and generalized expectation consistent-signal recovery (GEC-SR) \cite{8006946} algorithms, which are compressed sensing algorithms widely adopted in the mmWave channel estimation domain \cite{8171203, 8310593, 8320852}. In addition, we also adopt the DS-OMP channel estimator as our baseline, which is an algorithm that exploits the double sparsity structure of the UE-IRS-BS link by taking into account the fact that all the users share a common IRS-BS link. DS-OMP attempts to improve the channel estimation accuracy by identifying the common rows and columns that the UE-IRS-BS links of all the users share.

The channel parameters in \eqref{fgh} are as follows. The bandwidth and path loss are $W=80$ MHz and
\setcounter{equation}{55}
\begin{equation}
\mathrm{PL}=\begin{cases}31.4+20\log_{10}d&\text{for LoS}\\42+29.2\log_{10}d&\text{for NLoS}\end{cases}
\end{equation}
in dB where $d$ is the distance of the link in meters. The Rician K-factors are $\kappa_{\mathrm{UI}, k}=\kappa_{\mathrm{IB}}=13.2$ dB for the UE-IRS and IRS-BS links in LoS and $\kappa_{\mathrm{UB}, k}=-\infty$ dB for the UE-BS link in NLoS. The number of NLoS paths is $L_{\mathrm{UI}, k}=L_{\mathrm{IB}}=L_{\mathrm{UB}, k}=4$ for all the links.

In addition, the system parameters are configured as specified in the Dense Urban-eMBB scenario in ITU-R M.2412-0 \cite{itu2017guidelines}. In particular, the transmit power is $P_{k}[t]=23$ dBm. Meanwhile, the noise figure of the base station and active sensors is $\mathrm{NF}=7$ dB. Therefore, the noise power is $\sigma_{\mathrm{B}}^{2}=\sigma_{\mathrm{I}}^{2}=W\times N_{0}\times\mathrm{NF}$ where $N_{0}=-174$ dBm/Hz is the noise spectral density. There are $M=16$ antennas at the base station, $N=8\times 8$ elements at the IRS with $N_{\mathrm{a}}=4$ semi-passive elements, and $K=4$ users. The parameters for the active sensors are $B=4$ bit for the ADC resolution and $f_{\mathrm{sn}}=1$ for the switching frequency. In addition, the uniform $B$-bit quantizer proposed in \cite{8171203, 1057548} is adopted. The passive reflecting elements are initially turned off for UE-BS link estimation and turned on as
\begin{equation}\label{v}
|v_{n}[t]|=\begin{cases}0&\text{for }t\in\llbracket 50\rrbracket\\1&\text{for }t\in\llbracket T\rrbracket\setminus\llbracket 50\rrbracket\end{cases}
\end{equation}
with random phase shifts in the channel estimation phase. The base station and IRS are at $(0, 0)$ m and $(20, 10)$ m, while the users are around the circle with center $(40, 0)$ m and radius 5 m. The lengths of the coherence block and channel estimation phase are $T_{\mathrm{c}}=1800$ and $T=400$. Throughout the simulation, the system parameters are fixed unless stated otherwise. In addition, the parameters for VI-SBL are $M_{\mathrm{g}}=M, N_{\mathrm{g}}=N$, $S_{\mathbf{f}}=1$, $S_{\mathbf{g}}=8$, $S_{\mathbf{h}}=1$, and $a=b=10^{-6}$. To understand the reason behind $S_{\mathbf{g}}=8$, recall that the complexity of VI-SBL is $\mathcal{O}(N_{\mathrm{g}}^{3}K^{3}/S_{\mathbf{f}}^{2}+M_{\mathrm{g}}^{3}N_{\mathrm{g}}^{3}/S_{\mathbf{g}}^{2}+M_{\mathrm{g}}^{3}K^{3}/S_{\mathbf{h}}^{2})$. Since massive MIMO and IRS imply large $M_{\mathrm{g}}\geq M$ and $N_{\mathrm{g}}\geq N$, most of the complexity comes from \eqref{mgi} that repeats an $M_{\mathrm{g}}N_{\mathrm{g}}/S_{\mathbf{g}}\times M_{\mathrm{g}}N_{\mathrm{g}}/S_{\mathbf{g}}$ matrix inversion $S_{\mathbf{g}}$ times. Therefore, we convert a large matrix inversion to many small matrix inversions by setting $S_{\mathbf{g}}=8$.

\begin{table}[t]
\centering
\caption{Power consumption parameters}\label{table_1}
\begin{tabular}{|c|c|c|c|}
\hline
\textbf{Parameter}&\textbf{Value}&\textbf{Parameter}&\textbf{Value}\\
\hline
$B_{\mathrm{\infty}}$&10 bit&$P_{\mathrm{T}}$ \cite{8891922}&0.25 W/Gbps\\
\hline
$\mathrm{FOM}$ \cite{murmann2021adc}&1432.1 fJ/conversion-step&$P_{\mathrm{LO}}$ \cite{8333733}&22.5 mW\\
\hline
$f_{\mathrm{s}}$&80 MHz&$P_{\mathrm{RF}}$ \cite{8333733}&31.6 mW\\
\hline
\end{tabular}
\end{table}

Now, we explain the power consumption model that accounts for the power dissipated at the base station, IRS, and fronthaul link through which the received signal at the active sensors is forwarded. Since the base station is equipped with a local oscillator, $M$ RF chains, and $M$ pairs of $B_{\mathrm{\infty}}$-bit ADCs where $B_{\mathrm{\infty}}\gg 1$, the power consumed by the base station is modeled as $P_{\mathrm{BS}}=P_{\mathrm{LO}}+M(P_{\mathrm{RF}}+2P_{\mathrm{ADC}}(B_{\infty}))$. Here, $P_{\mathrm{ADC}}(B)=\mathrm{FOM}\times f_{\mathrm{s}}\times 2^{B}$ \cite{4403893} is the power consumption of a $B$-bit ADC where $\mathrm{FOM}$ and $f_{\mathrm{s}}$ are the figure of merit and sampling frequency. Likewise, the power consumed by the active sensors at the IRS is modeled as $P_{\mathrm{IRS}}=P_{\mathrm{LO}}+N_{\mathrm{a}}(P_{\mathrm{RF}}+2P_{\mathrm{ADC}}(B))$. Moving on to the fronthaul link that forwards $2BN_{\mathrm{a}}W$ bits per second, the power consumption model is $P_{\mathrm{FH}}=2BN_{\mathrm{a}}WP_{\mathrm{T}}$ \cite{8891922} where $P_{\mathrm{T}}$ is the traffic-dependent power consumption. Then, the total power consumption is $P_{\mathrm{BS}}+T/T_{\mathrm{c}}\times(P_{\mathrm{IRS}}+P_{\mathrm{FH}})$ under the premise that the semi-passive elements become passive reflecting elements in the data transmission phase. On the other hand, the total power consumption of the baselines with passive reflecting elements is $P_{\mathrm{BS}}$. The power consumption parameters are shown in Table \ref{table_1}.

\begin{table}[t]
\centering
\caption{Per-iteration complexities of various channel estimators}\label{table_2}
\begin{tabular}{|c|c|}
\hline
\textbf{Algorithm}&\textbf{Complexity}\\
\hline
Proposed&$\mathcal{O}(N_{\mathrm{g}}^{3}K^{3}/S_{\mathbf{f}}^{2}+M_{\mathrm{g}}^{3}N_{\mathrm{g}}^{3}/S_{\mathbf{g}}^{2}+M_{\mathrm{g}}^{3}K^{3}/S_{\mathbf{h}}^{2})$\\
\hline
GAMP&$\mathcal{O}(MTM_{\mathrm{g}}N_{\mathrm{g}}K)$\\
\hline
VAMP&$\mathcal{O}(MTM_{\mathrm{g}}N_{\mathrm{g}}K)$\\
\hline
SBL&$\mathcal{O}(M_{\mathrm{g}}^{3}N_{\mathrm{g}}^{3}K^{3})$\\
\hline
GEC-SR&$\mathcal{O}(M_{\mathrm{g}}^{3}N_{\mathrm{g}}^{3}K^{3})$\\
\hline
DS-OMP&$\mathcal{O}(MTK+NTKL_{\mathrm{IB}}L_{\mathrm{UI}, k}^{3})$\\
\hline
\end{tabular}
\end{table}

\begin{figure}[t]
\centering
\includegraphics[width=1\columnwidth]{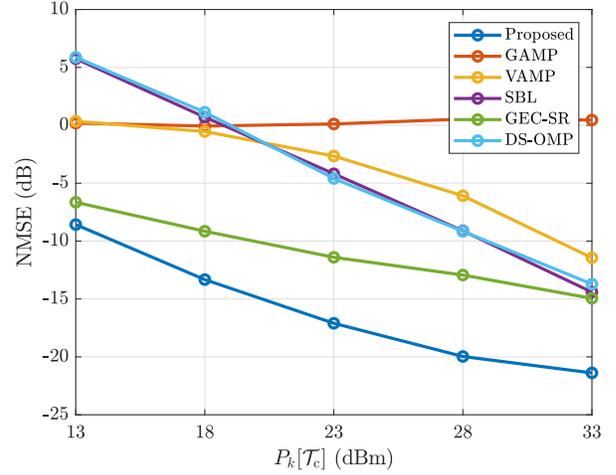}
\caption[caption]{NMSE of the UE-IRS-BS link vs. transmit power in the channel estimation phase.}\label{figure_3}
\end{figure}

\begin{figure}[t]
\centering
\includegraphics[width=1\columnwidth]{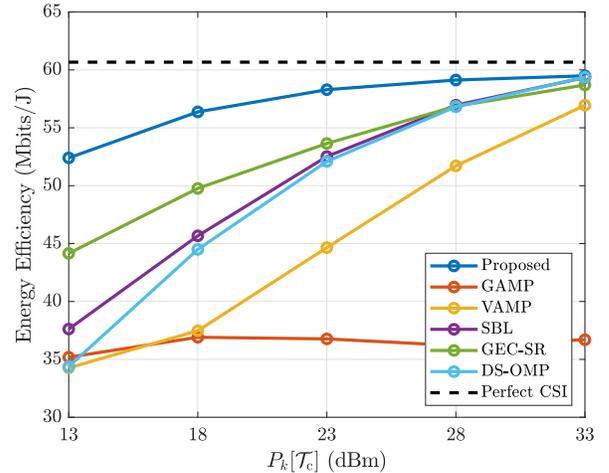}
\caption[caption]{Energy efficiency vs. transmit power in the channel estimation phase. The transmit power in the data transmission phase is $P_{k}[\mathcal{T}_{\mathrm{d}}]=23$ dBm.}\label{figure_4}
\end{figure}

The performance metrics for the channel estimation error and energy efficiency are as follows. The channel estimation error is measured based on the normalized MSE (NMSE). In particular, the NMSEs for the UE-IRS, IRS-BS, and UE-BS links are
\begin{align}
\mathrm{NMSE}(\bar{\mathbf{F}})&=\mathbb{E}\{\|\hat{\bar{\mathbf{F}}}-\bar{\mathbf{F}}\|_{\mathrm{F}}^{2}/\|\bar{\mathbf{F}}\|_{\mathrm{F}}^{2}\},\\
\mathrm{NMSE}(\bar{\mathbf{G}})&=\mathbb{E}\{\|\hat{\bar{\mathbf{G}}}-\bar{\mathbf{G}}\|_{\mathrm{F}}^{2}/\|\bar{\mathbf{G}}\|_{\mathrm{F}}^{2}\},\\ \mathrm{NMSE}(\bar{\mathbf{H}})&=\mathbb{E}\{\|\hat{\bar{\mathbf{H}}}-\bar{\mathbf{H}}\|_{\mathrm{F}}^{2}/\|\bar{\mathbf{H}}\|_{\mathrm{F}}^{2}\}.
\end{align}
Meanwhile, the spectral efficiency is evaluated based on the sum rate obtained by optimizing the passive reflecting elements \cite{8855810} and combiner at the base station \cite{9390351} using the estimates of the UE-IRS-BS link $\mathrm{kr}(\bar{\mathbf{F}}^{\mathrm{T}}, \bar{\mathbf{G}})$ and UE-BS link. Then, the energy efficiency is defined as the spectral efficiency scaled by $W$ and normalized by the total power consumption, which is defined as such to make a fair comparison between IRS with semi-passive elements and passive reflecting elements. Therefore, the energy efficiency is given by
\begin{equation}
\mathrm{EE}=\begin{cases}\frac{W\times\mathrm{SE}}{P_{\mathrm{BS}}+T/T_{\mathrm{c}}\times(P_{\mathrm{IRS}}+P_{\mathrm{FH}})}&\text{for }N_{\mathrm{a}}>0\\\frac{W\times\mathrm{SE}}{P_{\mathrm{BS}}}&\text{for }N_{\mathrm{a}}=0\end{cases}
\end{equation}
where the first case corresponds to VI-SBL with semi-passive elements, while the second case corresponds to the baselines with passive reflecting elements. Throughout the simulation, we mainly compare the NMSE of the UE-IRS-BS link because the baselines can only estimate the UE-IRS-BS and UE-BS links unlike VI-SBL that estimates all the links. Moreover, the reason for omitting the UE-BS link is because the UE-BS link estimation problem can be converted to the conventional channel estimation problem as pointed out in \cite{9130088}.

\begin{figure}[t]
\centering
\includegraphics[width=1\columnwidth]{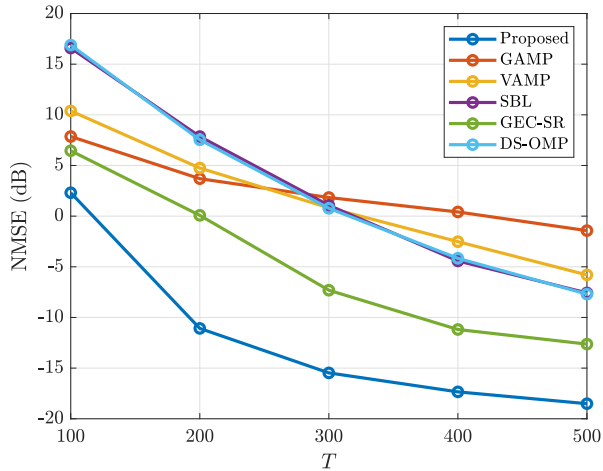}
\caption[caption]{NMSE of the UE-IRS-BS link vs. $T$.}\label{figure_5}
\end{figure}

\begin{figure}[t]
\centering
\includegraphics[width=1\columnwidth]{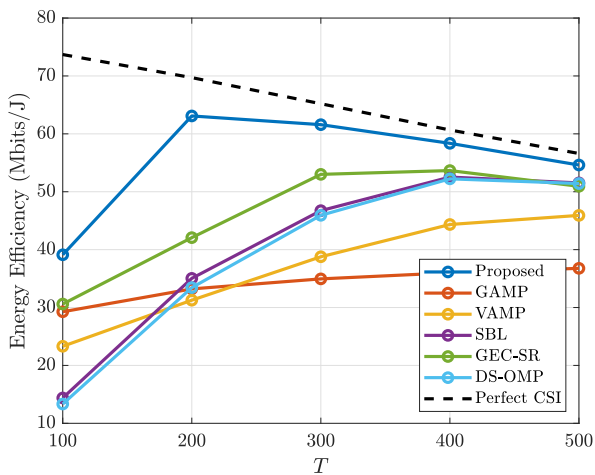}
\caption[caption]{Energy efficiency vs. $T$.}\label{figure_6}
\end{figure}

\begin{figure}[t]
\centering
\includegraphics[width=1\columnwidth]{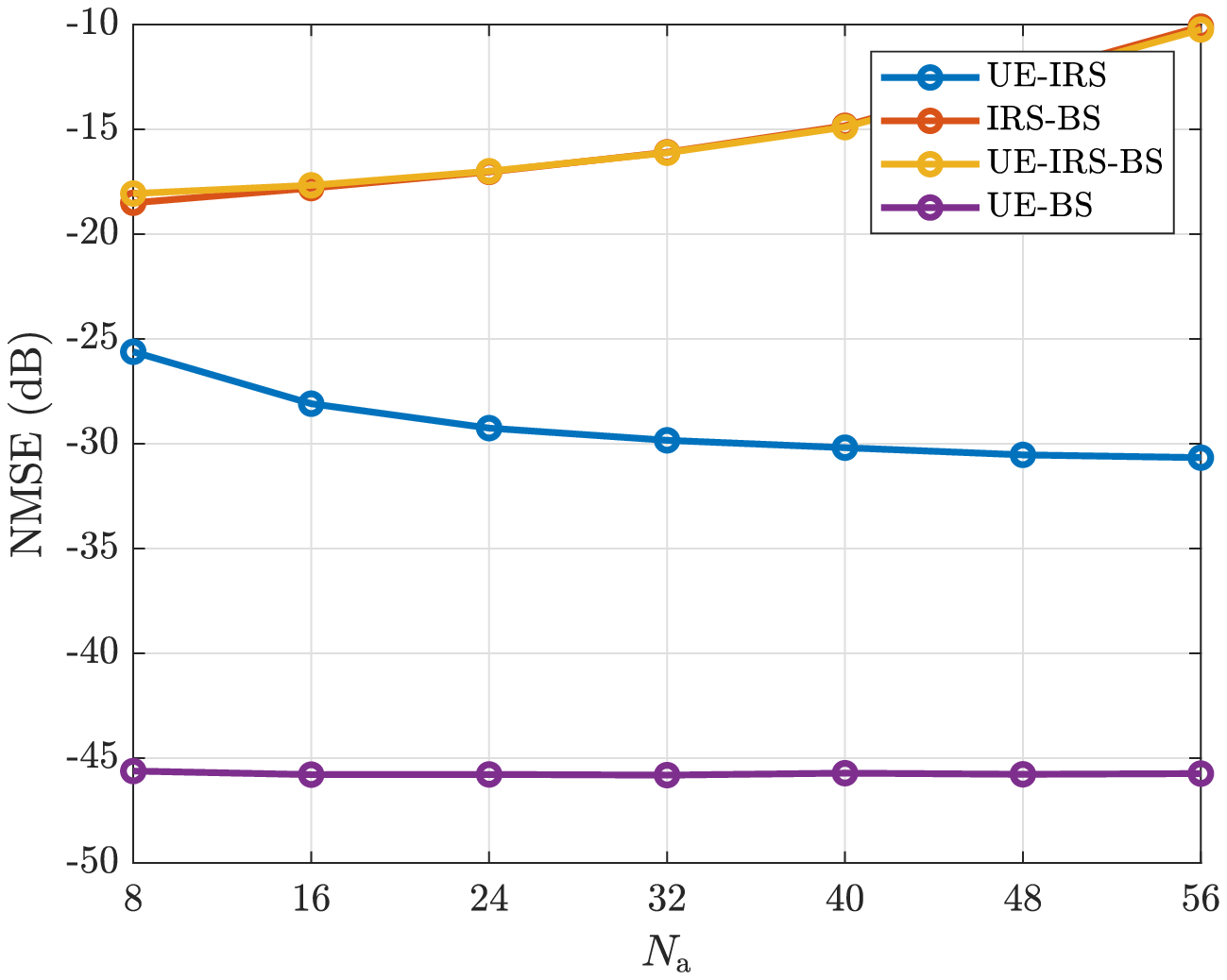}
\caption[caption]{NMSEs of all the links vs. $N_{\mathrm{a}}$. The baselines are not shown.}\label{figure_7}
\end{figure}

\begin{figure}[t]
\centering
\includegraphics[width=1\columnwidth]{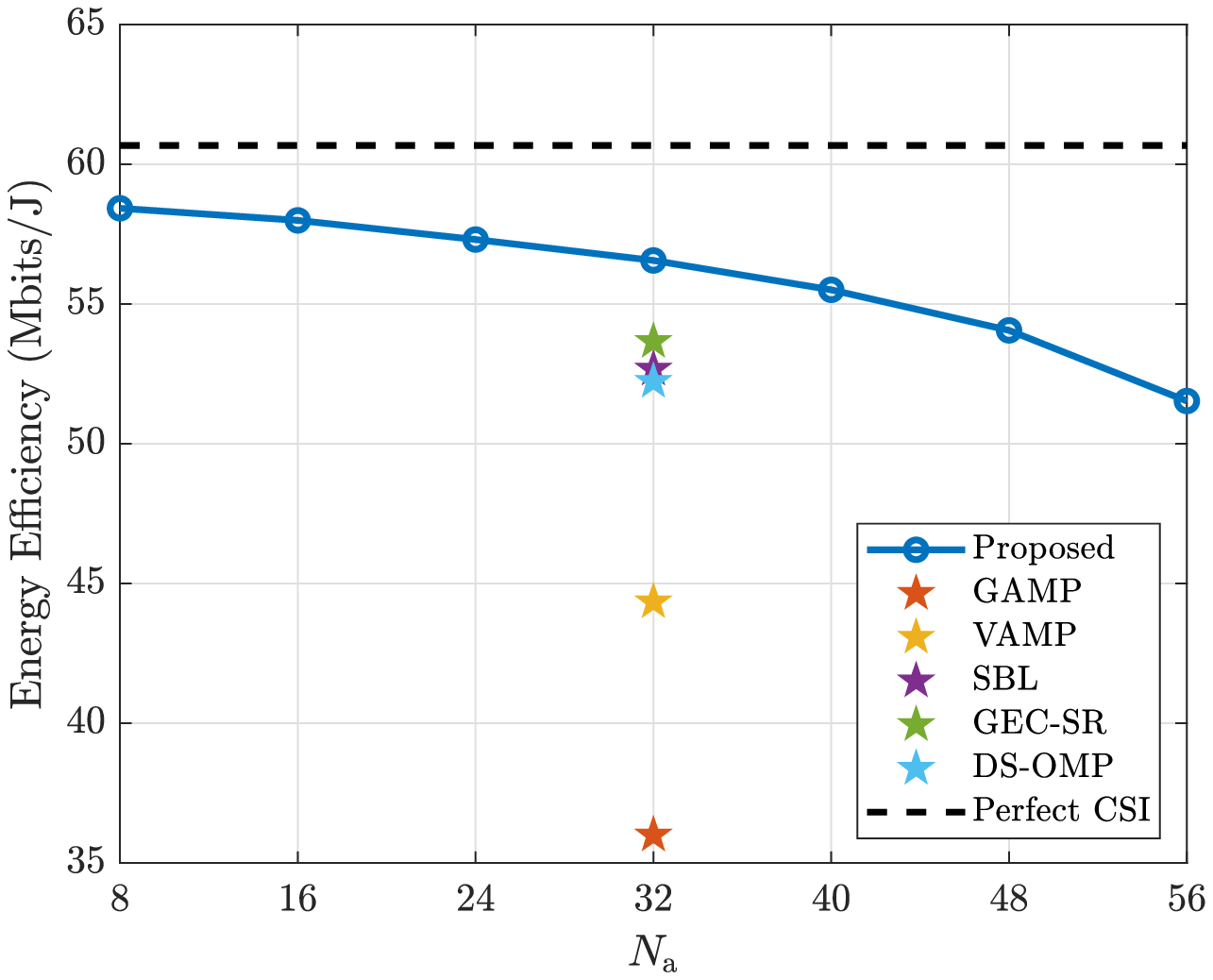}
\caption[caption]{Energy efficiency vs. $N_{\mathrm{a}}$. The baselines with passive reflecting elements are shown as references.}\label{figure_8}
\end{figure}

\begin{figure}[t]
\centering
\includegraphics[width=1\columnwidth]{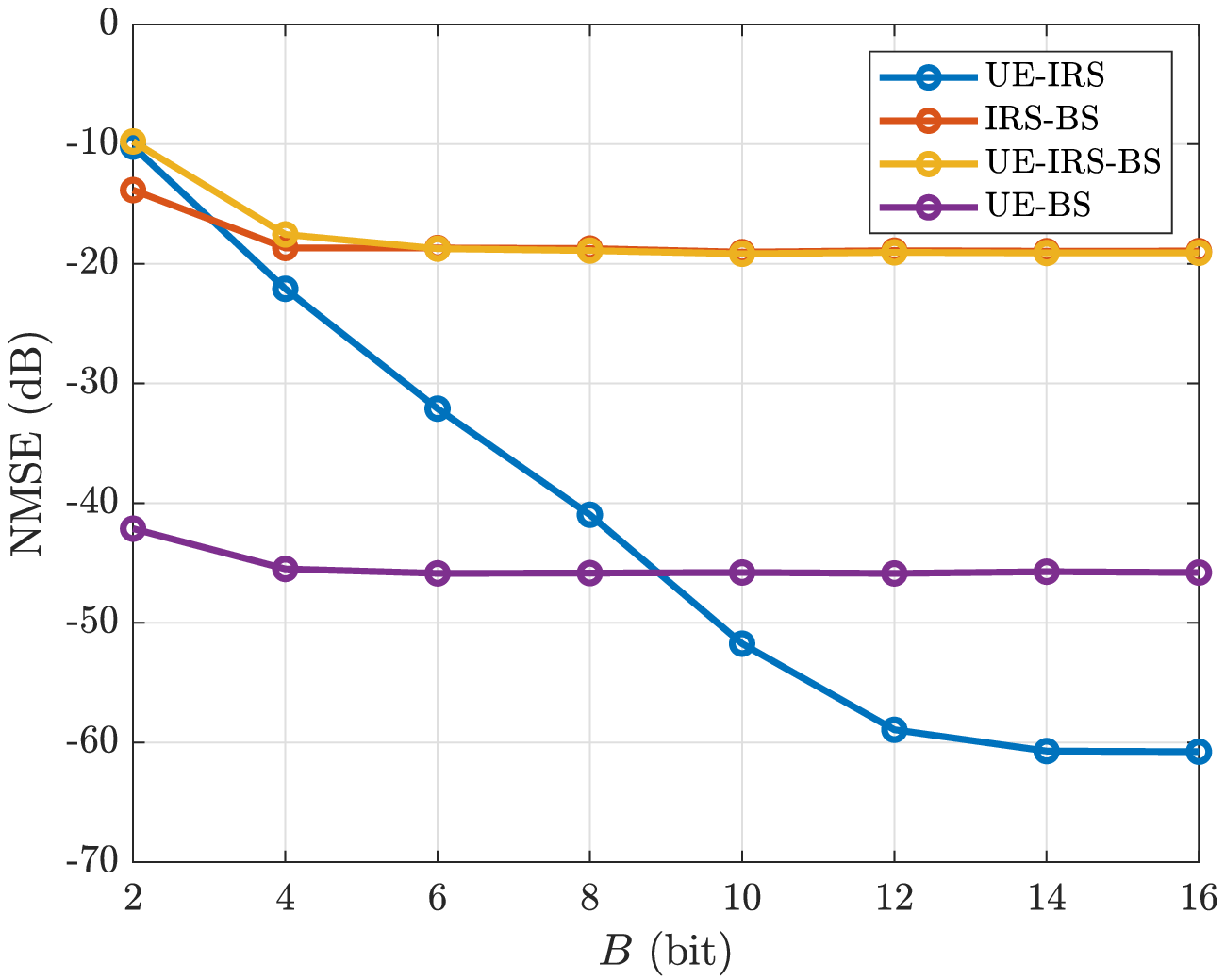}
\caption[caption]{NMSEs of all the links vs. $B$. The baselines are not shown.}\label{figure_9}
\end{figure}

\begin{figure}[t]
\centering
\includegraphics[width=1\columnwidth]{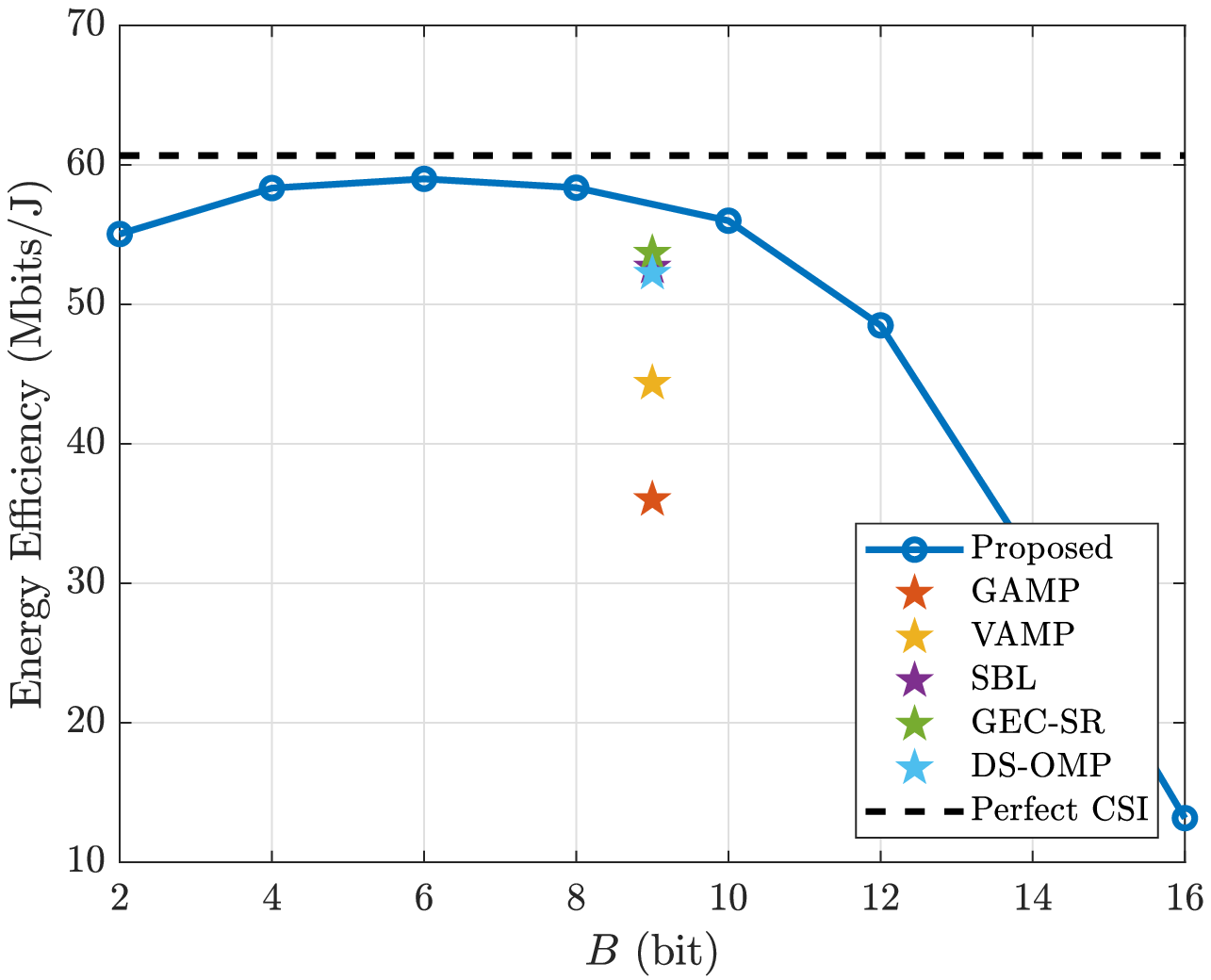}
\caption[caption]{Energy efficiency vs. $B$. The baselines with passive reflecting elements are shown as references.}\label{figure_10}
\end{figure}

\begin{figure}[t]
\centering
\includegraphics[width=1\columnwidth]{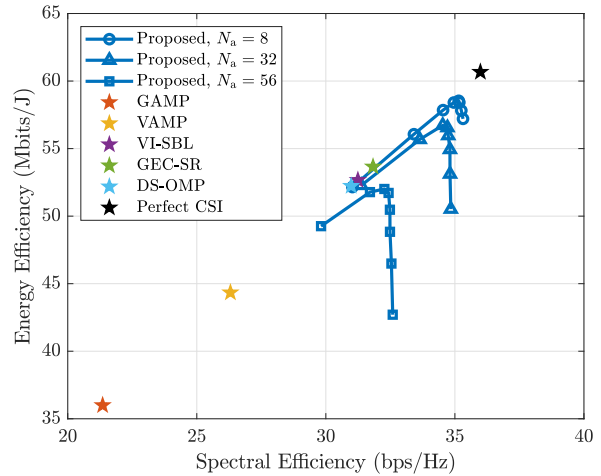}
\caption[caption]{Energy-spectral efficiency as a function of $B$ for various $N_{\mathrm{a}}$. At each line, the markers correspond to $B=1, \dots, 8$ from left to right. The baselines with passive reflecting elements are shown as references.}\label{figure_11}
\end{figure}

In the first simulation, we compare the performance of VI-SBL and the baselines in terms of the NMSE of the UE-IRS-BS link and energy efficiency for various transmit powers in the channel estimation phase. According to Fig. \ref{figure_3}, VI-SBL outperforms the baselines in terms of the channel estimation accuracy. The superior performance of VI-SBL is expected because VI-SBL exploits the additional information acquired from a small number of active sensors at the IRS, whereas only the received signal at the base station is available to the baselines. As a result, we see a significant energy efficiency gap from Fig. \ref{figure_4} due to the high channel estimation accuracy of VI-SBL that outweighs the additional power consumption.

In the second simulation, we evaluate the performance of various channel estimators for different $T$. According to Fig. \ref{figure_5}, $T\geq 200$ is sufficient for VI-SBL to yield an accurate channel estimate, whereas the baselines require at least $T\geq 400$. As a result, the energy efficiency gap is large in the low training overhead regime as evident from Fig. \ref{figure_6}. The reason for the dramatic training overhead gap comes from the fact that the baselines cannot observe the UE-IRS link, whereas VI-SBL has direct access to the UE-IRS link through a small number of active sensors. Therefore, the training overhead issue problematic in IRS-aided mmWave massive MIMO systems incurred by the large size of the UE-IRS-BS link can be resolved by adopting a small number of active sensors at the cost of additional but marginal power consumption.

In the third simulation, we evaluate the performance of VI-SBL in terms of the NMSEs of all the links and energy efficiency for various $N_{\mathrm{a}}$. From Fig. \ref{figure_7}, observe that the NMSE of the UE-IRS link decreases as $N_{\mathrm{a}}$ increases, which is not as surprising. In contrast, the NMSE of the IRS-BS link increases unlike the UE-IRS link. The reason for such a phenomenon is because the IRS-BS link can only be observed through the reflected signal. Therefore, more active sensors means less passive reflecting elements, or equivalently less IRS-BS link measurements. In addition, another interesting point is that the NMSE of the UE-IRS-BS link composed of the UE-IRS and IRS-BS links is lower-bounded by the worst NMSE of the UE-IRS and IRS-BS links as evident from Fig. \ref{figure_7}. From the discussion until now, we recommend to keep $N_{\mathrm{a}}$ small because increasing $N_{\mathrm{a}}$ results in additional power consumption, while degrading the quality of the UE-IRS-BS link estimate. By recalling that the impact of the UE-IRS-BS link on the spectral efficiency is significant, we conclude that increasing $N_{\mathrm{a}}$ offers no benefit other than increasing the accuracy of the UE-IRS link estimate. In fact, the energy efficiency falls below the baselines as $N_{\mathrm{a}}$ increases as evident from Fig. \ref{figure_8}.

In the fourth simulation, we assess the performance of VI-SBL in terms of the NMSEs of all the links and energy efficiency for various $B$. According to Fig. \ref{figure_9}, the NMSEs of all the links except the UE-IRS link saturate at $B\geq 4$. In contrast, the NMSE of the UE-IRS link hits the floor at $B\geq 12$. To understand such a phenomenon, recall that the UE-IRS link is observed through the quantized received signal at the active sensors. Then, we conclude that the impact of $B$ on the quality of the UE-IRS link estimate must be significant. Also, another interesting point is that the NMSE of the UE-IRS-BS link is lower-bounded by the worst NMSE of the UE-IRS and IRS-BS links as previously observed in Fig. \ref{figure_7}. Moving on to Fig. \ref{figure_10}, we see that there is a diminishing return beyond $B\geq 6$ at the cost of additional power consumption. Therefore, we can find the most energy-efficient operating point as a function of $B$, which is $B=6$ in our simulation setup.

In the fifth simulation, we investigate the energy-spectral efficiency of VI-SBL as a function of $B$ for various $N_{\mathrm{a}}$. According to Fig. \ref{figure_11}, it is evident that increasing the ADC resolution beyond a certain limit, i.e., approximately $B\geq 4$ for any $N_{\mathrm{a}}$, yields only a small spectral efficiency gain at the cost of large additional power consumption. Therefore, it is not worth it to deploy high-resolution ADCs at the active sensors of the IRS since the gain is marginal. In addition, an interesting observation is that increasing the number of active sensors at the IRS actually decreases the spectral efficiency. Therefore, the combined effect of reduced spectral efficiency and increased power consumption results in significantly degraded system performance. This is in line with Figs. \ref{figure_7} and \ref{figure_8}, and this phenomenon can be attributed to the fact that increasing the number of active sensors decreases the amount of signals reflected at the IRS towards the base station, thus decreasing the IRS-BS link measurements. Therefore, we conclude that an energy-efficient architecture that attains high energy efficiency requires both the number of the active sensors and ADC resolution to be low.

Before moving on, we emphasize that the superior performance of VI-SBL with active sensors over the baselines with passive reflecting elements is due to the availability of the additional information acquired at the active sensors. That is, the simulation results until now do not imply that VI-SBL proposed in this paper is algorithmically superior over the existing compressed sensing algorithms. The main contribution of the proposed VI-SBL is that it suggests a systematic way of how to jointly process the pilots received at the base station and active sensors effectively to yield good channel estimation accuracy/energy efficiency. This is in contrast to the recent works on IRS with semi-passive elements \cite{9511813, 9529045} that only use the pilots received at the active sensors due to the complicated nature of jointly exploiting the pilots received at the base station and active sensors. The channel estimators in \cite{9511813, 9529045} indeed require a dedicated channel estimator protocol involving both uplink and downlink pilots, which make them to be only applicable to limited scenarios.

Next, we investigate the complexities of various channel estimators. According to Table \ref{table_2}, SBL and GEC-SR that perform an $M_{\mathrm{g}}N_{\mathrm{g}}K\times M_{\mathrm{g}}N_{\mathrm{g}}K$ matrix inversion have the highest complexities. In contrast, GAMP and VAMP have relatively low complexities that involve a matrix-vector multiplication of size $MT\times M_{\mathrm{g}}N_{\mathrm{g}}K$, while DS-OMP has the lowest complexity slightly larger than that of the conventional OMP algorithm \cite{4385788}. Meanwhile, VI-SBL performs many small matrix inversions, whose complexity is controlled by $S_{\mathbf{f}}$, $S_{\mathbf{g}}$, and $S_{\mathbf{h}}$ as discussed earlier. The complexity of VI-SBL is demanding but not as SBL and GEC-SR. Therefore, we conclude that VI-SBL offers a significant performance gain at the expense of not-so-low complexity.

\section{Conclusion}\label{section_5}
A Bayesian channel estimator was proposed for IRS-aided mmWave massive MIMO systems with semi-passive elements. Unlike recent works on channel estimation with semi-passive elements that require both uplink and downlink signals to estimate the UE-IRS and IRS-BS links, the proposed channel estimator aimed to estimate all the links using only uplink training signals. To perform approximate posterior inference on the channel, the channel estimation problem was recast as an SBL framework. Then, SBL was solved using the variational free energy principle and mean-field approximation, from which the VI-SBL-based channel estimator was obtained. The simulation results showed that the proposed channel estimator requires low training overhead by taking advantage of the active sensors. Also, VI-SBL was capable of estimating all the links, which reduces the channel estimation overhead in the long run by replacing UE-IRS-BS link estimation with UE-IRS link estimation as the IRS-BS link is quasi-static in practice.

\appendices

\begin{figure*}[t]
\begin{align}
\langle\mathbf{A}_{\mathbf{gf}}^{\mathrm{H}}\mathbf{A}_{\mathbf{gf}}\rangle&\overset{\phantom{(a)}}{=}\langle(\mathbf{A}_{\mathrm{I}}^{\mathrm{T}}(\mathbf{S}^{*}\odot\mathbf{A}_{\mathrm{I}}^{*}\mathbf{F}^{*}\mathbf{X}^{*})\otimes\mathbf{A}_{\mathrm{B}}^{\mathrm{H}})((\mathbf{S}^{\mathrm{T}}\odot\mathbf{X}^{\mathrm{T}}\mathbf{F}^{\mathrm{T}}\mathbf{A}_{\mathrm{I}}^{\mathrm{T}})\mathbf{A}_{\mathrm{I}}^{*}\otimes\mathbf{A}_{\mathrm{B}})\rangle\notag\\
                                                                           &\overset{(a)}{=}\langle\mathbf{A}_{\mathrm{I}}^{\mathrm{T}}(\mathbf{S}^{*}\odot\mathbf{A}_{\mathrm{I}}^{*}\mathbf{F}^{*}\mathbf{X}^{*})(\mathbf{S}^{\mathrm{T}}\odot\mathbf{X}^{\mathrm{T}}\mathbf{F}^{\mathrm{T}}\mathbf{A}_{\mathrm{I}}^{\mathrm{T}})\mathbf{A}_{\mathrm{I}}^{*}\otimes\mathbf{A}_{\mathrm{B}}^{\mathrm{H}}\mathbf{A}_{\mathrm{B}}\rangle\notag\\
                                                                           &\overset{\phantom{(a)}}{=}\mathbf{A}_{\mathrm{I}}^{\mathrm{T}}\langle(\mathbf{S}^{*}\odot\mathbf{A}_{\mathrm{I}}^{*}\mathbf{F}^{*}\mathbf{X}^{*})(\mathbf{S}^{\mathrm{T}}\odot\mathbf{X}^{\mathrm{T}}\mathbf{F}^{\mathrm{T}}\mathbf{A}_{\mathrm{I}}^{\mathrm{T}})\rangle\mathbf{A}_{\mathrm{I}}^{*}\otimes\mathbf{A}_{\mathrm{B}}^{\mathrm{H}}\mathbf{A}_{\mathrm{B}}\label{agfagf}
\end{align}
\hrulefill
\end{figure*}

\section{Derivation of $\langle\mathbf{A}_{\mathbf{fg}}^{\mathrm{H}}\mathbf{A}_{\mathbf{fg}}\rangle$ for Updating $q(\mathbf{f})$}\label{appendix_a}
As a preliminary, define $\mathbf{B}_{\mathbf{f}}$ from
\begin{equation*}
\mathbf{A}_{\mathbf{fg}}=(\mathbf{I}_{T}\otimes\mathbf{A}_{\mathrm{B}}\mathbf{G}\mathbf{A}_{\mathrm{I}}^{\mathrm{H}})\underbrace{\mathrm{diag}(\mathrm{vec}(\mathbf{S}))(\mathbf{X}^{\mathrm{T}}\otimes\mathbf{A}_{\mathrm{I}})}_{=\mathbf{B}_{\mathbf{f}}}
\end{equation*}
in \eqref{afg}. Then, expanding $\langle\mathbf{A}_{\mathbf{fg}}^{\mathrm{H}}\mathbf{A}_{\mathbf{fg}}\rangle$ leads to
\begin{align*}
\langle\mathbf{A}_{\mathbf{fg}}^{\mathrm{H}}\mathbf{A}_{\mathbf{fg}}\rangle&\overset{\phantom{(a)}}{=}\langle\mathbf{B}_{\mathbf{f}}^{\mathrm{H}}(\mathbf{I}_{\mathrm{T}}\otimes\mathbf{A}_{\mathrm{I}}\mathbf{G}^{\mathrm{H}}\mathbf{A}_{\mathrm{B}}^{\mathrm{H}})(\mathbf{I}_{T}\otimes\mathbf{A}_{\mathrm{B}}\mathbf{G}\mathbf{A}_{\mathrm{I}}^{\mathrm{H}})\mathbf{B}_{\mathbf{f}}\rangle\\
                                                                           &\overset{(a)}{=}\langle\mathbf{B}_{\mathbf{f}}^{\mathrm{H}}(\mathbf{I}_{T}\otimes\mathbf{A}_{\mathrm{I}}\mathbf{G}^{\mathrm{H}}\mathbf{A}_{\mathrm{B}}^{\mathrm{H}}\mathbf{A}_{\mathrm{B}}\mathbf{G}\mathbf{A}_{\mathrm{I}}^{\mathrm{H}})\mathbf{B}_{\mathbf{f}}\rangle\\
                                                                           &\overset{\phantom{(a)}}{=}\mathbf{B}_{\mathbf{f}}^{\mathrm{H}}(\mathbf{I}_{T}\otimes\mathbf{A}_{\mathrm{I}}\langle\mathbf{G}^{\mathrm{H}}\mathbf{A}_{\mathrm{B}}^{\mathrm{H}}\mathbf{A}_{\mathrm{B}}\mathbf{G}\rangle\mathbf{A}_{\mathrm{I}}^{\mathrm{H}})\mathbf{B}_{\mathbf{f}}
\end{align*}
where the mixed-product property of the Kronecker product was used in (a). Meanwhile, the vectorized version of $\langle\mathbf{G}^{\mathrm{H}}\mathbf{A}_{\mathrm{B}}^{\mathrm{H}}\mathbf{A}_{\mathrm{B}}\mathbf{G}\rangle$ in the middle of the last equality can be expressed as
\begin{equation*}
\mathrm{vec}(\langle\mathbf{G}^{\mathrm{H}}\mathbf{A}_{\mathrm{B}}^{\mathrm{H}}\mathbf{A}_{\mathrm{B}}\mathbf{G}\rangle)=\langle\mathbf{G}^{\mathrm{T}}\otimes\mathbf{G}^{\mathrm{H}}\rangle\mathrm{vec}(\mathbf{A}_{\mathrm{B}}^{\mathrm{H}}\mathbf{A}_{\mathrm{B}}).
\end{equation*}
Now, let us focus on the conjugate transpose of $\langle\mathbf{G}^{\mathrm{T}}\otimes\mathbf{G}^{\mathrm{H}}\rangle$, or equivalently $\langle\mathbf{G}^{*}\otimes\mathbf{G}\rangle$, which contains $M_{\mathrm{g}}\times N_{\mathrm{g}}$ submatrices of sizes $M_{\mathrm{g}}\times N_{\mathrm{g}}$. The $(i, j)$-th submatrix of $\langle\mathbf{G}^{*}\otimes\mathbf{G}\rangle$ can be vectorized as
\begin{equation*}
\mathrm{vec}(\langle g_{i+(j-1)M_{\mathrm{g}}}^{*}\mathbf{G}\rangle)=[\mathbf{C}_{\mathbf{g}}+\mathbf{m}_{\mathbf{g}}\mathbf{m}_{\mathbf{g}}^{\mathrm{H}}]_{:,i+(j-1)M_{\mathrm{g}}},
\end{equation*}
which can be reshaped and assembled to compute $\langle\mathbf{A}_{\mathbf{fg}}^{\mathrm{H}}\mathbf{A}_{\mathbf{fg}}\rangle$.

\section{Derivation of $\langle\mathbf{A}_{\mathbf{gf}}^{\mathrm{H}}\mathbf{A}_{\mathbf{gf}}\rangle$ for Updating $q(\mathbf{g})$}\label{appendix_b}
First, we expand $\langle\mathbf{A}_{\mathbf{gf}}^{\mathrm{H}}\mathbf{A}_{\mathbf{gf}}\rangle$ using \eqref{agf} to obtain \eqref{agfagf} at the top of the page where the mixed-product property of the Kronecker product was applied in (a). Now, let us focus on the transpose of $\langle(\mathbf{S}^{*}\odot\mathbf{A}_{\mathrm{I}}^{*}\mathbf{F}^{*}\mathbf{X}^{*})(\mathbf{S}^{\mathrm{T}}\odot\mathbf{X}^{\mathrm{T}}\mathbf{F}^{\mathrm{T}}\mathbf{A}_{\mathrm{I}}^{\mathrm{T}})\rangle$ in the middle of the last line of \eqref{agfagf}. To proceed, we define $\mathbf{B}_{\mathbf{g}}$ from
\begin{equation*}
\mathrm{vec}(\mathbf{S}\odot\mathbf{A}_{\mathrm{I}}\mathbf{F}\mathbf{X})=\underbrace{\mathrm{diag}(\mathrm{vec}(\mathbf{S}))(\mathbf{X}^{\mathrm{T}}\otimes\mathbf{\mathbf{A}_{\mathrm{I}}})}_{=\mathbf{B}_{\mathbf{g}}}\mathbf{f}.
\end{equation*}
Then, the $(i, j)$-th element of $\langle(\mathbf{S}\odot\mathbf{A}_{\mathrm{I}}\mathbf{F}\mathbf{X})(\mathbf{S}^{\mathrm{H}}\odot\mathbf{X}^{\mathrm{H}}\mathbf{F}^{\mathrm{H}}\mathbf{A}_{\mathrm{I}}^{\mathrm{H}})\rangle$ is
\begin{align*}
&[\langle(\mathbf{S}\odot\mathbf{A}_{\mathrm{I}}\mathbf{F}\mathbf{X})(\mathbf{S}^{\mathrm{H}}\odot\mathbf{X}^{\mathrm{H}}\mathbf{F}^{\mathrm{H}}\mathbf{A}_{\mathrm{I}}^{\mathrm{H}})\rangle]_{i, j}=\\
&\sum_{k=1}^{T}[\mathbf{B}_{\mathbf{g}}\langle\mathbf{f}\mathbf{f}^{\mathrm{H}}\rangle\mathbf{B}_{\mathbf{g}}^{\mathrm{H}}]_{i+(k-1)N, j+(k-1)N}=\\
&\sum_{k=1}^{T}[\mathbf{B}_{\mathbf{g}}(\mathbf{C}_{\mathbf{f}}+\mathbf{m}_{\mathbf{f}}\mathbf{m}_{\mathbf{f}}^{\mathrm{H}})\mathbf{B}_{\mathbf{g}}^{\mathrm{H}}]_{i+(k-1)N, j+(k-1)N},
\end{align*}
which can be assembled and rearranged to compute $\langle\mathbf{A}_{\mathbf{gf}}^{\mathrm{H}}\mathbf{A}_{\mathbf{gf}}\rangle$.

\bibliographystyle{IEEEtran}
\bibliography{refs_all}

\end{document}